\documentclass[conference]{IEEEtran}
\usepackage{fancyhdr}
\usepackage[normalem]{ulem}
\usepackage{amsmath}
\usepackage{multirow}
\usepackage{caption}
\usepackage{float}
\usepackage{amssymb}
\usepackage{mathtools}
\usepackage{xspace}
\usepackage{color}
\usepackage{comment}
\usepackage[flushleft]{threeparttable}
\usepackage[bottom]{footmisc}
\usepackage[noend]{algcompatible}
\usepackage{graphicx}
\usepackage{epstopdf}
\usepackage{pifont}
\usepackage{listings}
\usepackage{tikz}
\usepackage{subdepth}
\usepackage{subcaption}
\usepackage{algorithm}
\usepackage{dsfont}
\usepackage{color}
\usepackage[noend]{algpseudocode}

\newcommand{\bheading}[1]{{\vspace{2pt}\noindent{\textbf{#1}}\hspace{2pt}}}
\newcommand{\eheading}[1]{{\vspace{2pt}\noindent{\emph{#1}}\hspace{2pt}}}

\newcommand{\RNum}[1]{\uppercase\expandafter{\romannumeral #1\relax}}

\newcommand{\ie}{\emph{i.e.}\xspace}
\newcommand{\eg}{\emph{e.g.}\xspace}

\newcommand{\secref}[1]{\mbox{Sec.~\ref{#1}}\xspace}

\def\authnotes{1}
\newcommand{\authnote}[2]{\ifnum\authnotes=1\begin{quote}\textbf{#1 says:} #2\end{quote}\fi}
\newcommand{\fixme}[1]{\ifnum\authnotes=1\textbf{\textcolor{red}{[FIXME: #1]}}\fi}

\DeclareMathOperator*{\argmax}{argmax} 
\newcommand{\norm}[1]{\left\lVert#1\right\rVert}

\newenvironment{packeditemize}{
\begin{list}{$\bullet$}{
\setlength{\labelwidth}{8pt}
\setlength{\itemsep}{0pt}
\setlength{\leftmargin}{\labelwidth}
\addtolength{\leftmargin}{\labelsep}
\setlength{\parindent}{0pt}
\setlength{\listparindent}{\parindent}
\setlength{\parsep}{0pt}
\setlength{\topsep}{3pt}}}{\end{list}}

\author{\IEEEauthorblockN{Zecheng He}
\IEEEauthorblockA{Princeton University \\
Princeton, NJ \\
zechengh@princeton.edu}
\and
\IEEEauthorblockN{Tianwei Zhang}
\IEEEauthorblockA{Princeton University \\
Princeton, NJ \\
tianweiz@alumni.princeton.edu}
\and
\IEEEauthorblockN{Ruby B. Lee}
\IEEEauthorblockA{Princeton University \\
Princeton, NJ \\
rblee@princeton.edu
}
}

\begin{document}
\title{VerIDeep: Verifying Integrity of Deep Neural Networks through Sensitive-Sample Fingerprinting}

\maketitle
\begin{abstract}



Deep learning has become popular, and numerous cloud-based services are provided to help customers develop and deploy deep learning applications. Meanwhile, various attack techniques have also been discovered to stealthily compromise the model's integrity. When a cloud customer deploys a deep learning model in the cloud and serves it to end-users, it is important for him to be able to verify that the deployed model has not been tampered with, and the model's integrity is protected.

We propose a new low-cost and self-served methodology for customers to verify that the model deployed in the cloud is intact, while having only black-box access (e.g., via APIs) to the deployed model. Customers can detect arbitrary changes to their deep learning models. Specifically, we define \texttt{Sensitive-Sample} fingerprints, which are a small set of transformed inputs that make the model outputs sensitive to the model's parameters. Even small weight changes can be clearly reflected in the model outputs, and observed by the customer. Our experiments on different types of model integrity attacks show that we can detect model integrity breaches with high accuracy ($>$99\%) and low overhead ($<$10 black-box model accesses).



\end{abstract}

\section{Introduction}
\label{sec:intro}

The past few years have witnessed the fast development of deep learning (DL). One popular class of deep learning models is Deep Neural Networks (DNN), which has been widely adopted in many artificial intelligence applications, such as image recognition \cite{HeZhRe:15}, natural language processing \cite{LuPhMa:15} and speech recognition \cite{HaCaCa:14}. The state-of-the-art DNN models usually consist of a large number of layers and parameters. Therefore, it takes plenty of computational resources, storage and time to generate a model and deploy it in the product.

To make it automatic and convenient to deploy deep learning applications, many IT corporations offer cloud-based services for deep learning model training and serving, usually dubbed as Machine Learning as a Service (MLaaS). For model training, Google provides the Tensor Processing Unit (TPU) hardware \cite{googleTPU} to customers for machine learning acceleration. It also provides a Machine Learning Engine \cite{googleMLEngine} that enables customers to upload their Tensorflow code and training data to train models in the cloud. Microsoft offers Azure ML Studio \cite{microsoftazure} to achieve a similar task. Amazon provides the pre-built deep learning environment \cite{amazonami} and GPU instances \cite{amazonp2} to ease model training. For model serving, Amazon offers the SageMaker framework \cite{amazonsagemaker}, where customers provide models and release query APIs to end users to use the models.


Deploying deep learning tasks in MLaaS brings new security concerns. In particular, cloud customers may have concern about their model integrity when outsourcing the model for deployment in the cloud. First, an adversary can intentionally tamper with the model in the cloud, especially the one used by security-critical applications, to make it malfunction. For example, as deep learning based face recognition has become popular in access control and surveillance systems, the adversary may seek to modify the face classifier to bypass the access control or avoid being identified. Because a cloud environment is complex, involving a lot of entities and their interactions such as customers, network, cloud storage and ML applications, an adversary may exploit potential vulnerabilities in the network or storage protocols \cite{mulazzani2011dark, somorovsky2011all, bugiel2011amazonia} to get accesses to the target model in transit or at rest, and then modify the parameters to meet his attack goals. Different types of model integrity attacks have been proposed: (1) in a DNN trojan attack \cite{Trojannn, GuDoGa:17, ChLiLi:17}, the adversary can slightly modify the target DNN model to make it mis-classify the inputs containing a trigger predefined by the adversary, while classifying inputs without the trigger correctly; (2) in an error-generic data poisoning attack \cite{biggio2012poisoning, mei2015using, xiao2015feature}, the adversary can intentionally degrade the model accuracy of one specific class or the overall accuracy, via model fine-tuning on malicious training samples; the same technique can also be used in an error-specific attack \cite{munoz2017towards}, where the model mis-classifies a target class as an adversary desired class. These integrity breaches have caused significant security threats to DNN-based applications, such as autonomous driving \cite{GuDoGa:17, Trojannn} and user authentication \cite{ChLiLi:17}.


Second, a dishonest cloud provider may stealthily violate the Service Level Agreement (SLA), without making the customers aware, for financial benefits \cite{zhang2011homealone, BoVaJu:11}. For instance, in the context of deep learning based cloud service, the cloud provider can use a simpler or compressed model to replace the customer's models to save computational resources and storage \cite{GhGuGa:17}. It is easy to compress the model without being aware by the customer, since the customers need to grant the cloud providers access to the ML resources, \eg, training data or models, to use the cloud service. Customers are annoyed with such SLA violation, even though it has a subtle impact on the model accuracy, as they pay more for the resources than they actually get.

However, providing an approach to protecting model integrity of DNN models deployed in clouds is challenging. First, The complex cloud environment inevitably causes a big attack surface, and it is challenging to guarantee the integrity of the models during different cloud operations. Second, verifying the model integrity is very difficult. Once the customers submit their models to the clouds, the security status of the model is not transparent or directly verifiable to the customers. Traditional integrity attestation methods (\eg, calculating the hash values of protected data) can hardly work because the adversary or the dishonest cloud provider may provide unreliable attestation results to customers, making them believe the models are intact, when in fact they are not. Third, for some model integrity attacks, the adversary only makes subtle modifications to the model, and wrong predictions only occur for specific attacker-chosen inputs which are imperceptible to the customers. Therefore, it is difficult for the customers to verify the correctness of the models by checking the outputs. Fourth, the cloud provider may not actively check the data integrity status in a timely manner. It may be too late for the customers to realize the model corruption when he actually retrieves it. Therefore, providing a self-served and low-cost model verification approach brings a great benefit to the customer.

Past work has been designed to defeat model integrity attacks. For DNN trojan attacks, Liu et al. \cite{LiXiSr:17} proposed to detect anomalies in the dataset, or remove the trojan via model retraining or input preprocessing. For data poisoning attacks, the typical solution is also to identify and remove the poisoning data from the dataset by statistical comparisons \cite{ChStVa:17, LiLiVo:17, StKoLi:17}. While these methods are effective locally, they fail to protect the integrity of models in clouds, as the customers do not have the privileges to apply these protections in MLaaS, and the adversary can easily disable or evade these protection mechanisms. In the cloud scenario, Ghodsi \cite{GhGuGa:17} proposed a protocol to verify if an untrusted cloud provider cheats the customer with a simpler and less accurate model. However, this approach can hardly detect some subtle model integrity attacks which only slightly modify the models to make them behave correctly for normal inputs. 

In this paper, we propose a new methodology for customers to verify the integrity of deep learning models stored in the cloud, with only black-box access to the model. Specifically, we propose \texttt{Sensitive-Samples} as fingerprints of DNN models. \texttt{Sensitive-Samples} are minimally transformed inputs of the DNN model. They are carefully designed so that the model outputs of these samples are very sensitive to changes of model parameters. Even if the adversary makes only small changes to a small portion of the model parameters, the outputs of the \texttt{Sensitive-Samples} from the model also changes, which can be observed by the customer. Furthermore, these \texttt{Sensitive-Samples} are very similar to common inputs so the adversary cannot tell if they are used for normal model serving or integrity testing. Our methodology is generic and can be applied to different deep neural networks, with no assumptions about the network architecture, hyper-parameters, or training methods. Our experiments show that our proposed defense can detect all existing model integrity attacks with high detection rate ($>$99\%) and low cost ($<$10 black-box model accesses).

We show the feasibility of a new line of research where properties of a DNN model, such as the integrity of its parameters, can be dynamically checked by just querying the model with a few transformed inputs and observing their outputs. The key contributions of this paper are:

\begin{packeditemize}
\item A novel \texttt{Sensitive-Samples} generation approach for deep neural network integrity verification. It only needs black-box access to the cloud model through APIs, thus hard to be spotted by the attacker. Our proposed approach can be applied to general neural networks, with no assumptions on network architecture, hyper-parameters and training methods. 
\item A Maximum Active-Neuron Cover sample selection algorithm to generate the fingerprint of a DNN model from \texttt{Sensitive-Samples}, reduce the number of required \texttt{Sensitive-Samples} and further increase the detection capability.
\item Comprehensive evaluation of our approach on different types of integrity attacks on various applications and models. 
\end{packeditemize}

The rest of the paper is organized as follows: Section \ref{sec:bg} gives the background of deep learning, MLaaS, and the threat model. Section \ref{sec:ps} presents the problem statement. Section \ref{sec:method} describes our new methodology of detecting model integrity vulnerabilities. Section \ref{sec:implementation} introduces the detailed implementation, models and attack techniques for evaluation. Section \ref{sec:eval} describes the performance and security evaluations. Section \ref{sec:discussion} shows some discussions. We present related work in Section \ref{sec:related} and conclude in Section \ref{sec:conclu}.
\section{Background and Threat Model}
\label{sec:bg}

In this section, we describe the background of deep neural networks, machine learning as a service, and our threat model. 

\subsection{Deep Neural Networks}
\label{sec:bg:dl}

A deep neural network is a parameterized function $f_\theta: \mathcal{X} \mapsto \mathcal{Y}$ that maps an input $x\in \mathcal{X}$ to an output $y\in \mathcal{Y}$. Various neural network architectures have been proposed and applied to different tasks, \eg multilayer perceptrons \cite{Ro:58}, convolutional neural networks \cite{LeJaBo:89} and recurrent neural networks \cite{RuHiWi:86}.


A neural network usually consists of an input layer, an output layer and one or more hidden layers between the input and output. Each layer is a collection of units called \emph{neurons}, which are connected to other neurons in adjacent layers. Each connection between the neurons can transmit a signal to another neuron in the next layer. In this way, a neural network transforms the inputs through a sequence of hidden layers, and then the outputs by applying a linear function followed by an element-wise nonlinear activation function (\eg \texttt{sigmoid} or \texttt{ReLU}) in each layer, as shown in Eq (\ref{eq:mlp}).

{\small
\begin{align}
\label{eq:mlp}
\begin{split}
&h_1 = \phi_1(w_1x+b_1) \\
&h_2 = \phi_2(w_2h_1+b_2) \\
&\qquad \qquad ... \\
&h_n = \phi_n(w_nh_{n-1}+b_n) \\
&y = \texttt{softmax}(w_yh_n+b_y)
\end{split}
\end{align}
}%

The training process of a neural network is to find the optimal parameters $\theta$ that can accurately reflect the relationship between $\mathcal{X}$ and $\mathcal{Y}$. To achieve this, the user needs a training dataset $D^{train}=\{x^{train}_i, y^{train}_i\}^{N}_{i=1}$ with $N$ samples, where $x^{train}_i \in \mathcal{X}$ is the input and $y^{train}_i \in \mathcal{Y}$ is the corresponding ground-truth label. Then a loss function $L$ is adopted to measure the distance between the ground-truth output $y^{train}_i$ and the predicted output $f_\theta(x^{train}_i)$. The goal of training a neural network is to minimize this loss function (Eq (\ref{eq:loss})). Backward propagation \cite{GoBeCo:16} and stochastic gradient descent \cite{RoMo:51} are commonly used methods to achieve this goal. After figuring out the optimal parameters $\theta^*$, given a testing input $x^{test}$, we can easily predict the output $y^{test} = f_{\theta^*}(x^{test})$. This prediction process is called \texttt{inference}.
\begin{align}
\label{eq:loss}
\theta^* = \arg\min_{\theta}(\sum_{i=1}^N L(y^{train}_i, f_\theta(x^{train}_i))
\end{align}

\subsection{Machine Learning as a Service (MLaaS)}
\label{sec:bg:dldeployment}

Different frameworks and tools are designed to simplify the machine learning deployment, making machine learning more practical. Cloud-based services are released to provide automated solutions for data processing, model training, inference and further deployment. These services are called Machine Learning as a Service (MLaaS). Many public cloud providers have launched such machine learning services, \eg, Amazon Sagemaker \cite{amazonsagemaker}, Google Cloud ML Engine \cite{googleMLEngine}, Microsoft Azure ML Studio \cite{microsoftazure}. These services attract machine learning practitioners to deploy applications in the cloud without having to set up their own large-scale ML infrastructure.

\begin{figure}[t]
\centerline{\mbox{\includegraphics[width=0.85\linewidth]{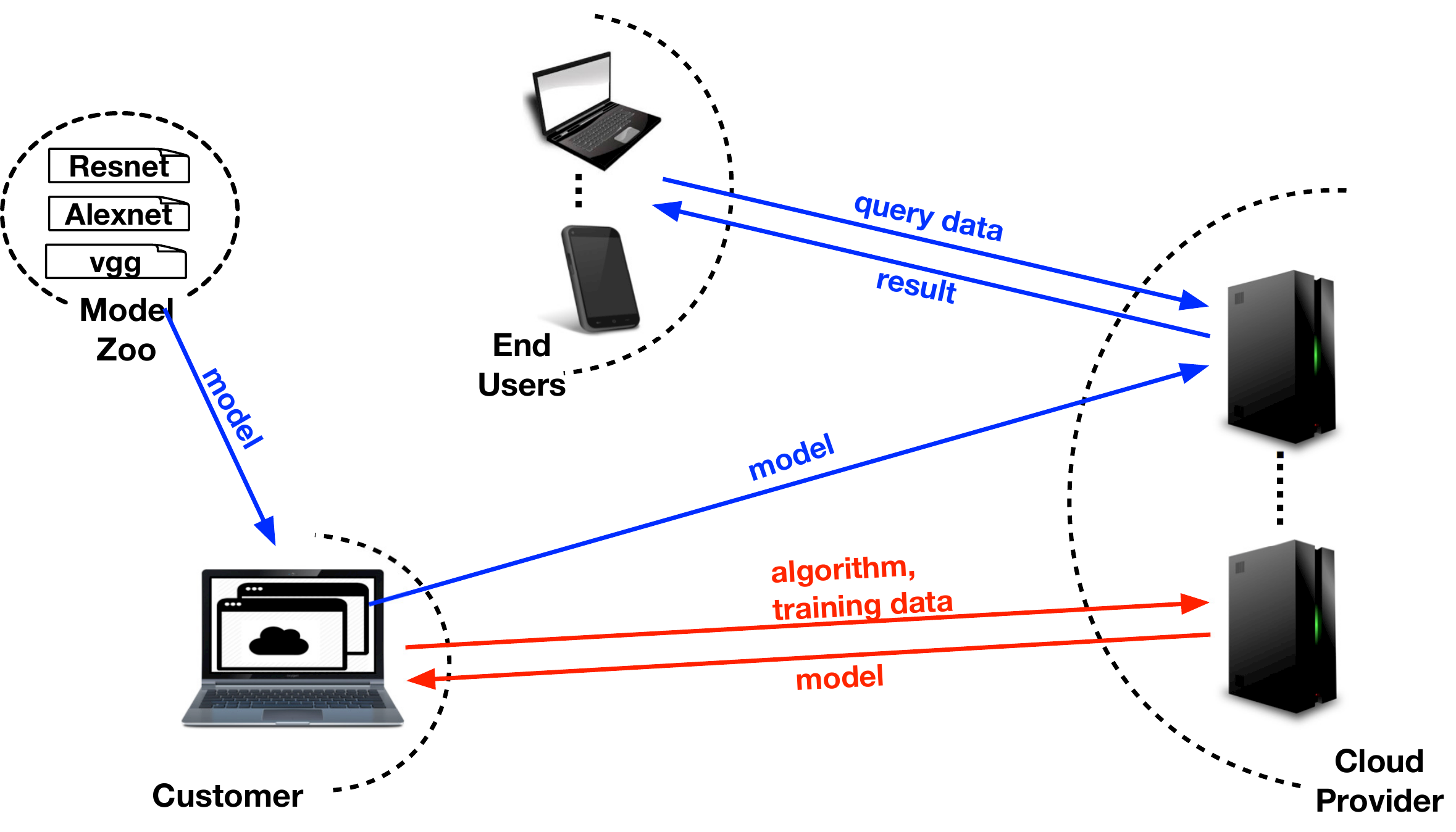}}}
\caption{Machine Learning as a Service. Red lines: the model \texttt{training} service. Blue lines: the model \texttt{serving} service.}
\label{fig:bg:mlaas}
\end{figure}

There are two main types of machine learning services in the cloud: model training and model serving. They are illustrated in Figure \ref{fig:bg:mlaas}. For model training, the cloud provider offers the computing resources (\eg, GPU, TPU) and machine learning environment (\eg, OS images, docker containers), and the customer selects a machine learning algorithm (it can be a default one from the cloud provider, or a customized one specified by the customer). Then the ML algorithm runs on the computing resources and generates the model for the customer. Customers are charged based on the duration and amount of resources used for their machine learning tasks.

For model serving, the customer uploads a model to the cloud platform. This model can be trained in the cloud, on the customer's local server, or downloaded from the public model zoo. The cloud platform allocates resources to host the model, sets up an endpoint, and releases query APIs to end users. Then the end users can use the ML model for prediction via these query APIs: they send input data to the endpoint, and the cloud platform conducts input pre-processing, model inference and output post-processing, and returns the final results to the end users. The cloud provider is responsible for service availability and scalability. Customers are charged on a pay-per-query basis. A cloud platform can offer both the model training and model serving services to customers to achieve end-to-end machine learning deployment. 

\section{Problem Statement}\label{sec:ps}
\subsection{Threat Model}

We consider the integrity of ML models in a cloud-based model \emph{serving} service. We do not consider the model \emph{training} service as it is not necessary for ML model deployment: instead of training the model in clouds, the customer can train it locally, or download it from a publicly verified model zoo. In another word, we make no assumptions about the model generation process.


We do not make specific assumptions about how the model integrity is compromised, and to what extent the model is modified. We consider, but not limited to, the following attacks to compromise the model integrity hosted in the cloud: \ding{172} An adversary can exploit the vulnerabilities of cloud network protocols or service interfaces \cite{somorovsky2011all} to tamper with the model when it is in transit between the customer and cloud provider; \ding{173} An adversary can also exploit the cloud storage vulnerabilities (e.g., \cite{mulazzani2011dark}) or OS images \cite{bugiel2011amazonia} to replace the model with a compromised one; \ding{174} The adversary can be the dishonest cloud provider who violates the SLA for financial benefits. For example, he can compress the model to save storage and computational resources. The dishonest cloud provider has full read/write access to the customer's ML models stored in the cloud. Traditional integrity attestation methods (e.g., calculating hash value of a protected model) can hardly work because the dishonest cloud provider may provide unreliable attestation results to customers.


We do not consider the integrity of DNN \emph{executions} by the cloud provider, since previous generic methods and systems have been proposed to protect the integrity of code execution in clouds, e.g. Intel SGX secure enclave \cite{mckeen2013innovative, anati2013innovative}. These work can be applied to the deep learning inference execution. Changing the parameters of the DL model is a more subtle attack, unique to DL, and no defense has been proposed before, which is the focus of this paper.





\subsection{Problem Statement}

We consider a problem, where a customer wants to verify the integrity of his model stored and served in the remote cloud. Formally, a customer $\mathcal{C}$ owns a deep neural network model $y=f_\theta(x)$, where $\theta$ is the set of all parameters in the model. This function $f$ takes a vector $x$ as input\footnote{If $x$ is a tensor, we can always vectorize it.} and outputs a vector $y$. The customer uploads the model $f_{\theta}$ to the cloud provider $\mathcal{P}$. The cloud provider sets up an endpoint, accepting new inputs $x$ from end users (as well as the customer), and returning the outputs of the model $y$. 

However, after the customer sends out the target model, the integrity of this model can be compromised by adversaries when it is in network transit, or in cloud storage. A dishonest cloud provider can also breach the model's integrity, e.g. compression. So the actual model served by the cloud provider (denoted as $f_{\theta'}$) may be different from the one uploaded by the customer ($f_{\theta}$). The customer's goal is to verify if $f_{\theta'}$ is the same as $f_{\theta}$. The customer has direct access to the reference model $f_{\theta}$ locally, before uploading it to the cloud, to generate our proposed fingerprint. This is the same assumption as using hashing for integrity protection. But he has only black-box access to the model $f_{\theta'}$: he can neither get the model parameters $\theta'$ nor take a hash; he can only send arbitrary inputs $x^*$ to the model and receive the outputs $y^*=f_{\theta'}(x^*)$. He needs to verify the model integrity based on the model input and output pairs $(x^*, y^*)$, and the correct model $f_\theta$ as the reference. 

\subsection{Use Cases}
\label{sec:ps:usecase}
We provide some use cases to illustrate the necessity of model integrity verification. 

\begin{figure}[h]
\centering
\includegraphics[width=0.3\textwidth]{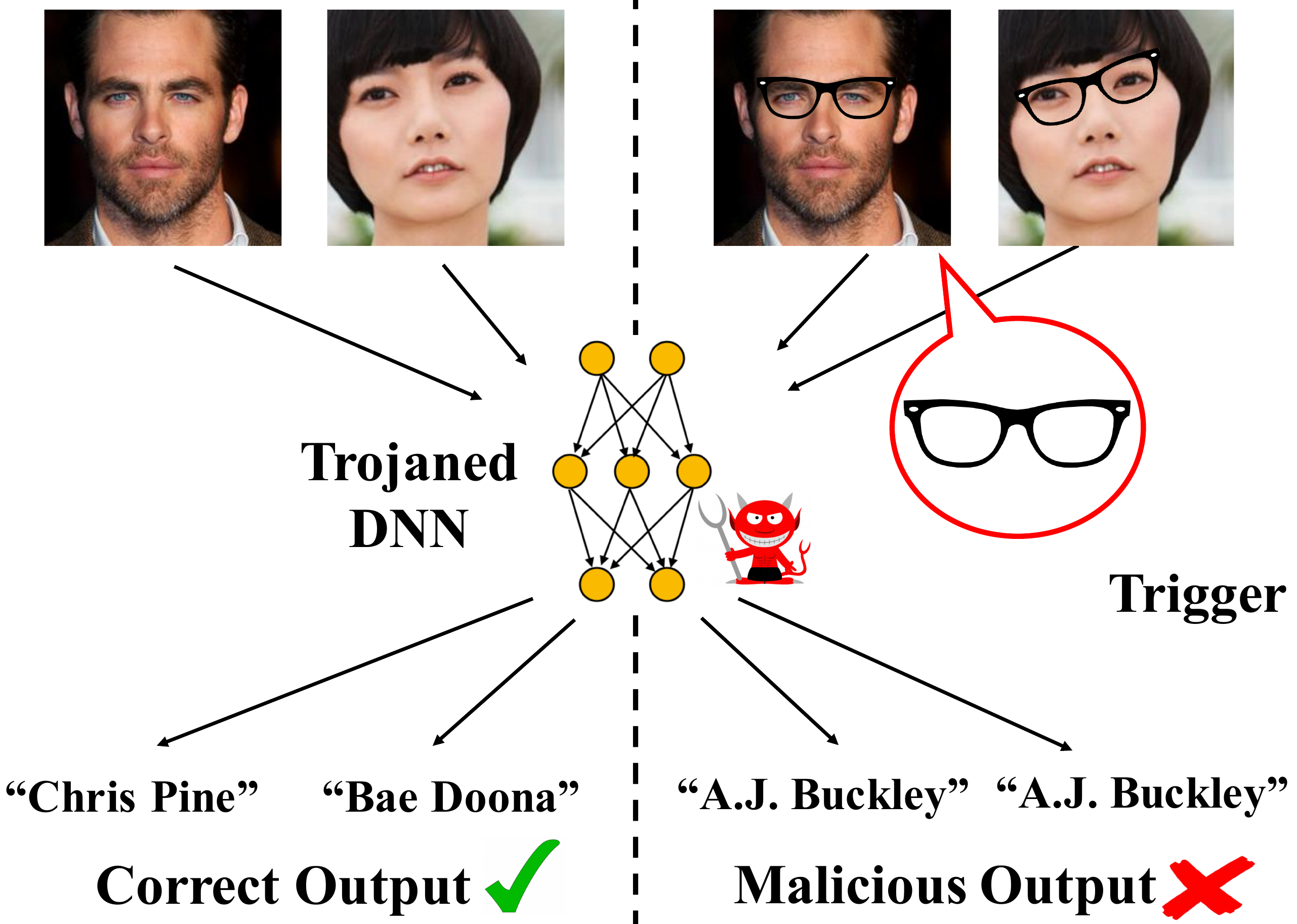}
\caption{Illustration of a DNN trojan. A person without the trigger (left) is recognized correctly by the trojaned DNN. A person wearing a specific pair of glasses, i.e. the trigger, is mis-classified.}
\label{fig:trojan-examples}
\end{figure}

\bheading{DNN trojan attacks}. Past work proposes DNN model trojan attacks \cite{Trojannn, geigel2013neural, geigel2014unsupervised, GuDoGa:17, ChLiLi:17, zou2018potrojan}, in which an adversary injects trojan into the model by slightly modifying the parameters from $\theta$ to $\theta'$. Then the compromised model still has the state-of-the-art performance on the end-user's validation and testing samples, i.e., when the end user sends a normal input sample $x$ to the model for prediction, the compromised model gives the same result as the correct model: $f_{\theta'}(x) = f_\theta(x)$. However, the compromised model behaves maliciously on specific attacker-chosen inputs. For instance, if the input sample contains a trigger $\Delta$, e.g. a pair of glasses in face recognition system \cite{ChLiLi:17}, then the compromised model gives a label $y'$ that does not match the output from the correct model, \ie, $y' = f_{\theta'}(x+\Delta) \neq f_\theta(x+\Delta)$. Note that $y'$ can be a fixed label pre-determined by the attacker, or an arbitrary unmatched label. We show an illustration of a DNN trojan in Figure \ref{fig:trojan-examples}. The customer trains a DNN model for face authentication. When the adversary injects trojan into the DNN model, the original faces can be correctly classified. However, if the person wears a specific pair of glasses, the trojaned model will always predict the image as ``A. J. Buckley''. The adversary can use this technique to bypass the authentication mechanism.

The adversary has multiple ways to inject the trojan into the ML models. He can incrementally retrain the model with selected critical neurons and weights \cite{Trojannn}, fine-tune the model from a correct one with poisoned data \cite{GuDoGa:17}, or even train the model from scratch with the poisoned dataset \cite{ChLiLi:17, geigel2013neural, geigel2014unsupervised}. Then in the MLaaS scenario, the adversary can be a man-in-the-middle attacker who replaces the model with a trojaned one during the model transit from the customers to the cloud. The adversary can also exploit the cloud storage vulnerabilities \cite{mulazzani2011dark} to inject the trojan into the model when it has already been stored in clouds.

\bheading{Targeted data poisoning attacks\footnote{Indiscriminate attack, i.e. degrading the accuracy of all samples, can also be done through data poisoning. However, such modification can be trivially detected by taking normal samples as inputs and examining the outputs. Therefore, we do not explicitly discuss the indiscriminate attack in the paper.}.} In this type of attacks, the adversary tries to mis-classify a specific set of samples. The attack can be error-specific, if the sample is mis-classified as a specific class; or error-generic, if the sample is mis-classified as any of the incorrect classes. For instance, in an autonomous-driving task, a customer trains a DNN model for traffic sign recognition. The adversary can compromise the model in a way that the model always predicts the ``STOP'' sign as other traffic signs (\eg, ``SPEED LIMIT 60'', ``NO STOPPING ANYTIME''), while maintaining the accuracy of other signs. Then this compromised model can cause severe traffic accidents if the compromised model is in production.

Modifying the model to mis-classify a specific set of samples can be achieved by data poisoning: the adversary fine-tunes the original correct one, with the augmented dataset consisting of carefully designed malicious samples. Then similar to DNN trojan attacks, the adversary can replace the original model with the new compromised one from the network, storage or cloud provider in MLaaS operations. Detecting such model modification via only black-box access is non-trivial, because the customer does not know the attacker's target set.

\bheading{Storage saving attacks}. 
The cloud provider may compromise the model integrity in an innocuous way driven by financial incentive. A dishonest cloud provider may replace the customer's model with a simpler one to save computational resources during model inference \cite{GhGuGa:17}. He can also use different techniques (e.g., pruning \cite{HaMaJ:16}, quantization \cite{GoLiYa:14}, low precision \cite{CoBeDa:14}, etc) to compress customers' models to save storage. For instance, the cloud provider can use these techniques to compress a VGG-16 model for image classification, saving the storage by $49\times$ with the accuracy degradation of only $0.33\%$ \cite{HaMaJ:16}. Therefore, it is very hard for customers to detect if this model is compromised with only black-box access to the target model. 

\subsection{Challenges}

Protecting the integrity of DNN models in remote clouds is a challenging task for customers. First, the adversary has different means to tamper with the models in the complex MLaaS cloud environment. It is beyond the customers' capability to guarantee or check the model integrity after the model leaves the customers.

Second, once the targeted model is uploaded, the customers do not have direct access to the model files in the cloud. So they cannot directly compare the model parameters (or hashes) with the correct ones and check if they are intact. Even if the MLaaS has APIs to enable customers to retrieve the models or their hash values from the cloud side, a dishonest cloud provider can give customers the original models, cheating them that their models are correct, when a compromised model is actually served.

Third, it is possible that the customer exhaustedly sends inputs to the cloud, retrieves the label and compares it with the reference outputs. Ghodsi et al. \cite{GhGuGa:17} used such a method to verify if an untrusted cloud provider cheats the customer by serving a simpler and less accurate model. But it is very inefficient to detect the attacks in \secref{sec:ps:usecase}, because the attacker explores the minimal influence on ``non-target'' inputs to avoid being detected in these attacks. Ideally, a trojaned model gives the wrong prediction \textit{only when the input contains the trigger}. Since the triggers are unknown to the customer, the customer has no means to produce a sample, whose outputs from the correct model and trojaned model are distinguishable. For targeted data poisoning attacks, the target set is unknown to the customer. For storage saving attacks, model compression does hardly affect the model output results, so the output of normal samples cannot reflect the model integrity.

\section{Sensitive-Sample Fingerprinting}
\label{sec:method}

\subsection{Overview of Our Methodology}
\label{sec:method:mainidea}
As described in \secref{sec:ps}, we consider the attack scenario in which the customer uploads a machine learning model $f_\theta$ to the cloud provider for model serving. However, an adversary may compromise the model and change it to $f_{\theta'}$. The customer wants to verify if the model served by the cloud provider is actually the one he uploaded. The model served by the cloud provider is a black-box to the customer, \ie the customer does not have direct access to the model files in the cloud. He can only use the online model for prediction: when he sends an input to the cloud provider, the cloud provider returns the corresponding output. 

\textbf{To deal with the online black-box scenario, our main idea is that, we can generate a small set of transformed inputs $\{v_i\}_{i=1}^n$, whose outputs predicted by the compromised model will be different from the outputs predicted by the original intact model.} We call such transformed inputs \texttt{Sensitive-Samples}. We use a small set of these transformed inputs and their corresponding correct model outputs as the \textbf{\textit{fingerprint}} of the DNN model, i.e. $\mathbb{FG}=\{(v_i, f_\theta(v_i))\}_{i=1}^n$.

To verify the integrity of a model, the customer first uses the correct model locally to generate \texttt{Sensitive-Samples} and obtain the corresponding correct output $y = f_\theta(v)$. Then he simply sends these samples to the cloud provider and obtains the output $y' = f_{\theta'}(v)$. By comparing $y$ and $y'$, the customer can check if the model is intact or changed. 



There are some challenges in designing a good fingerprint, especially a good input transform, for integrity checking. We define a qualified fingerprint as one satisfying the following characteristics:

\begin{packeditemize}

\item \textbf{Sensitive to model changes.} In order to detect the model integrity breach, the fingerprint must be sensitive to the modification of model parameters. In some attacks, the adversary changes a small number of parameters, e.g. selective neuron modification to inject trojans or backdoors \cite{Trojannn}. We must carefully design the input transform so that they are able to capture model changes made by real attacks.



\item \textbf{Generalizable.} The fingerprint generation algorithm should be independent of the machine learning models and the training datasets. The same method should be generalizable to different models and datasets.

\item \textbf{Hard to be spotted.} The fingerprint cannot be a random or weird input. Otherwise, the adversary can easily notice the unusual usage. Instead, the generated fingerprint should look similar to natural inputs so the adversary cannot recognize if it is used for integrity checking, or for normal model serving.

\item \textbf{Easy to verify.} The integrity checking algorithm must be as simple as possible, in order to reduce the cost and overhead for the verification.

\end{packeditemize}


To achieve the above requirements, we convert the generation of Sensitive-Samples to an optimization problem: the transformed inputs must make the model outputs the most sensitive to the parameters under such inputs. So even when a small portion of the model parameters are changed (as in the backdoor or trojan attacks), the corresponding outputs of these samples change with the highest probability. 
We generate the \texttt{Sensitive-Samples} from common inputs so they are quite similar to the common ones and can hardly be spotted by the adversary. The \texttt{Sensitive-Samples} can detect the integrity breaches while the common inputs cannot. Below we first formally define the problem of generating \texttt{Sensitive-Samples} in Section \ref{sec:method:PS}. Then we detail the algorithm of generating samples in Section \ref{sec:method:gen}, and fingerprints in Section \ref{sec:method:MANC}. We discuss the effects of model output specification on the sample generation in Section \ref{sec:method:outputspec}.

\subsection{Integrity Verification Goal} \label{sec:method:PS}
A DNN model can be defined as a function $y=f_\theta(x)$. Here $\theta$ is the set of all parameters in the model. 
We rewrite the model function as $y=f(W,x)=[y_1,...,y_r]^{T}=[f_1(W,x),...,f_r(W,x)]^{T}$. Here $W=[w_1, w_2,...,w_s]$ is a subset of parameters-of-interest in $\theta$ in our consideration, containing the weights and biases.


We assume $W$ in the correct model is modified by $\Delta w$, i.e. $W'=W+\Delta w$. The corresponding outputs of the correct and compromised model become $y=f(W, x)$ and $y'=f(W+\Delta w, x)$, respectively. In order to precisely detect this change through $y$ and $y'$, the ``sensitive'' input $v$ should maximize the difference between $y$ and $y'$. Formally, this means:

\begin{align}
\begin{split}
v &= argmax_x \ \ ||f(W+\Delta w, x)-f(W, x)||_2 \\
&= argmax_x \ \ ||f(W+\Delta w, x)-f(W, x)||_2^2 \\
&= argmax_x\ \ \Sigma_{i=1}^{r} ||f_i(W+\Delta w, x)-f_i(W, x)||_2^2
\end{split}
\end{align}

\noindent where $||\cdot||_2$ denotes the $l2$ norm of a vector. Note that we do not have prior-knowledge on $\Delta w$ (how the adversary modifies the model) in advance. With Taylor Expansion we have:
\begin{align}
f_i(W+\Delta w,x) &= f_i(W,x)+\frac{\partial f_i(W,x)}{\partial W}^{T}\Delta w+O(||\Delta w||_{2}^{2}) \label{eq:Taylor}
\end{align}

Consider $\Delta w$ as a perturbation of $W$, we approximate Eq (\ref{eq:Taylor}) to the first-order term:
\begin{align}
||f_i(W+\Delta w,x)-f_i(W,x)||_2^2 &\approx || \frac{\partial f_i(W,x)}{\partial W}^T\Delta w||_2^2 \label{eq:diff} \\
&\propto ||\frac{\partial f_i(W,X)}{\partial W}||_2^2 \label{eq:propto}
\end{align}

Note that the left-hand side of Eq (\ref{eq:diff}) models the difference of output $y_i$ between a correct DNN and a compromised DNN.



In Eq (\ref{eq:propto}) we conclude that the $l2$ norm of the gradient $||\frac{\partial f_i(W,x)}{\partial W}||_2$ can model the element-wise ``sensitivity'' of the DNN output corresponding to the parameters. Therefore, the sensitivity $S$ of $f(W,x)$ can be defined as:

\begin{align}
\begin{split}
S &= \Sigma_{i=1}^{r} ||\frac{\partial f_i(W,x)}{\partial W}||_{2}^2 \\
&={\norm{\frac{\partial f(W,x)}{\partial W}}}^{2}_{F} \label{eq:sensitivity}
\end{split}
\end{align}
where $|| \cdot ||_{F}$ is the Frobenius norm \cite{Fnorm} of a matrix.

We would like to look for the optimal input $v$ to maximize the sensitivity $S$ through the following optimization problem:
\begin{align}
v = \argmax_{x} {\norm{\frac{\partial f(W,x)}{\partial W}}}^{2}_{F}
\end{align}

A \texttt{Sensitive-Sample} of the model is the transformed input and the corresponding output, i.e. $(v,y=f(W,v))$.

\subsection{Sensitive Sample Generation} \label{sec:method:gen}

Generating sensitive samples is an optimization problem. We introduce some constraints on the samples. 

\bheading{Sample Correctness.}
In some cases, there are some requirements for the range of sample data, denoted as $[p, q]$. For instance, in the case of image-based applications, the intensities of all pixels must be in the range of [0, 255] for a valid image input. 

\bheading{Small Perturbation.}
In Section \ref{sec:method:mainidea}, we described a \texttt{sensitive-sample} should look like a normal input, to prevent the adversary from evading the integrity checking. So we add one more constraint: the generated sample is a small perturbation of a natural data $v_0$ sampled from original data distribution $\mathcal{D}_X$, i.e. the difference of the generated sample and $v_0$ should not exceed a small threshold $\epsilon$.

Eqs (\ref{eq:object}) summarize the objective and constraints of this optimization problem. The constraint set $[p,q]^m$ is a convex set, therefore we can use Projected Gradient Ascent \cite{projectedgd} to generate $v$.
\begin{align}
\label{eq:object}
\begin{split}
v =& \argmax_{x} {\norm{\frac{\partial f(W,x)}{\partial W}}}^{2}_{F}\\
&\text{s.t. } x\in[p,q]^m\\
&\norm{x-v_0}\le \epsilon
\end{split}
\end{align}

We show our proposed \texttt{Sensitive-Sample} generation algorithm in Algorithm \ref{alg:gen}. Lines 9-10 initialize the input, i.e. a sample from the natural data distribution $\mathcal{D}_X$. Line 12 sets up the element-wise loss function $||\frac{\partial f_i(W,x)}{\partial W}||_{2}^2$. Line 13 sets up the sample correctness constraints. Line 14 loops while $v$ is still similar to the original initialization $v_0$. $itr\_max$ is set to avoid an infinite loops. Lines 15-19 apply a gradient ascent on the sensitivity, a.k.a. S in Eq (\ref{eq:sensitivity}).  Line 20 projects $v$ onto the sample correctness constraint set. It is a truncation of the values to the range of [p,q], which is [0,255] for image pixels.

\begin{algorithm}[h]
\caption{Generating a \texttt{Sensitive-Sample}}\label{alg:gen}
\begin{algorithmic}[1]
\State \textbf{Function} Sensitive-Sample-Gen($f$, $W$, itr\_max, $\epsilon$, lr)
\\
\State /* f: the target model */
\State /* W: parameters in consideration */
\State /* itr\_max: maximum number of iterations */
\State /* $\epsilon$: threshold for small perturbation constraints */
\State /* lr: learning rate in projected gradient ascent */
\\
\State $v_0 $= Init$\_$Sample()
\State $v = v_0$
\State $i = 0$
\State $l_k = {\norm{\frac{\partial f_k(W,v)}{\partial W}|}}^{2}_{2}$, \ \ $k={1,2...N_{Output}}$
\State Constraint$\_$Set = $[p,q]^m$
\While{(($|v-v_0|\le \epsilon$) \&\& ($i<itr\_max$))}
    \State $\Delta = 0$
    \For{($k=0; k<N_{Output}; k++$)}
        \State $\Delta += \partial l_k / \partial v$
    \EndFor \State \textbf{end for}
    \State $v = v + lr * \Delta$
    \State $v$=Projection($v$, Constraint$\_$Set)
    \State $i++$
\EndWhile \State \textbf{end while}
\State \textbf{return} $\{v,f(W, v)\}$
\end{algorithmic}
\end{algorithm}

\subsection{Fingerprint Generation: Maximum Active-Neuron Cover (MANC) Sample Selection} \label{sec:method:MANC}

In some cases, one \texttt{Sensitive-Sample} may not be enough to detect all possible model changes. The main reason is, we observe that if a neuron is inactive\footnote{The neuron's output after the activation function is 0 or very close to 0.} given an input sample, the sensitivity of all weights connected to that neuron becomes zeros, i.e. the small modification of such weights will not be reflected in the outputs.

\begin{figure}[h]
\centering
\includegraphics[width=0.16\textwidth]{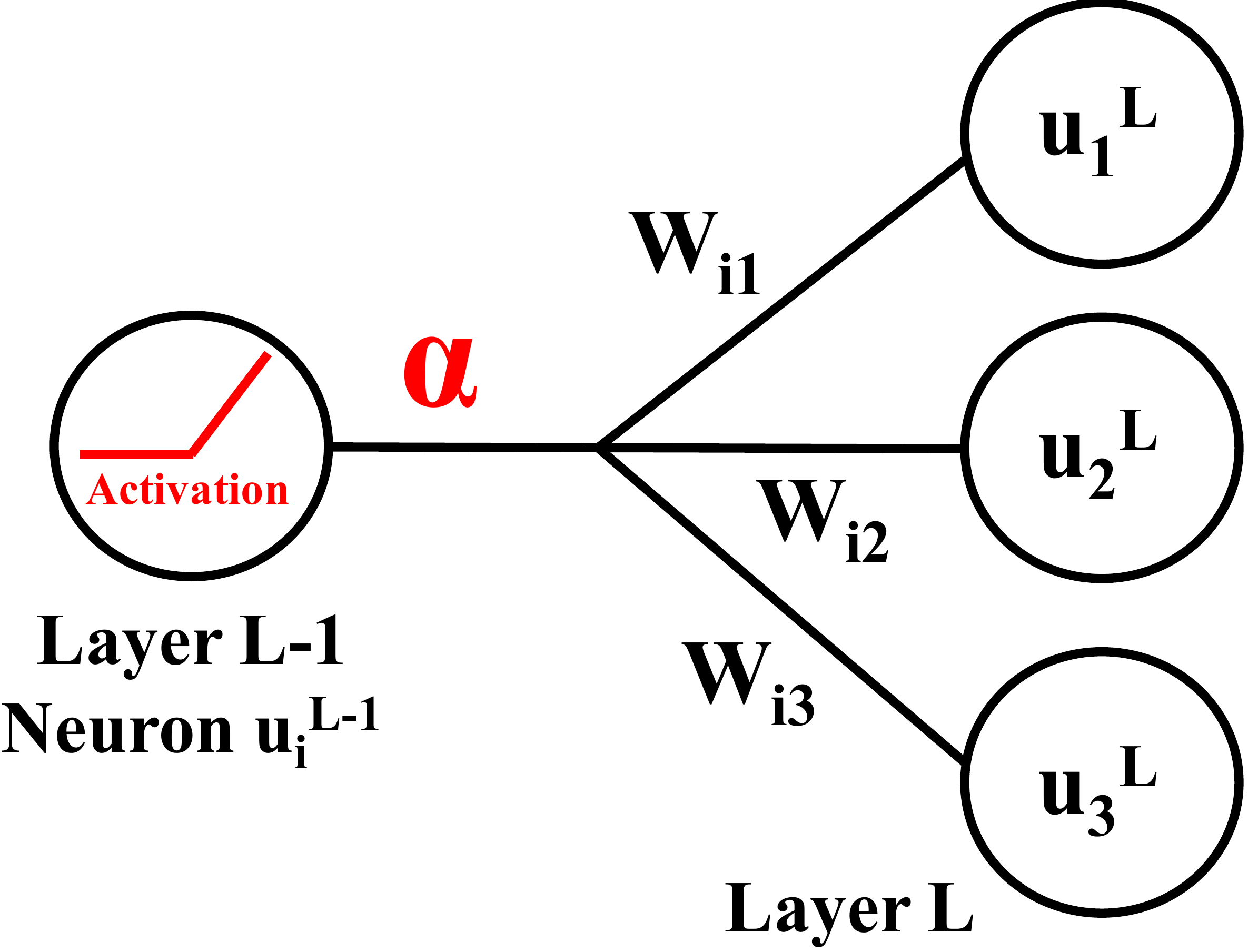}
\caption{Backward propagation of DNN. Inactive neuron $u_{i}^{L-1}$ zeros the sensitivity of weights connected to it ($W_{i1},W_{i2},W_{i3}$).}
\label{fig:inactive-neuron}
\end{figure}

Figure \ref{fig:inactive-neuron} illustrates this scenario. Suppose the activation function of layer $L-1$ is \texttt{ReLU}\footnote{If the activation function is sigmoid, a neuron is inactive if $\alpha$ is very close to 0.}, and the activation $\alpha$ of a neuron $u_i^{L-1}$ in the previous layer $L-1$ is 0 (inactive). Based on the chain-rule of backward propagation, we have:

\begin{align}
\begin{split}
\frac{\partial f}{\partial w_{i1}} &= \frac{\partial f}{\partial u_{1}^{L}} \frac{\partial u_{1}^{L}}{\partial w_{i1}} \\
&= \frac{\partial f}{\partial u_{1}^{L}} \alpha \\
&= 0
\end{split}
\end{align}

Thus by Taylor Expansion we have:

\begin{align}
\label{eq:nodiff}
\begin{split}
f(w_{i1}+\Delta w_{i1},x) &= f(w_{i1},x)+\frac{\partial f}{\partial w_{i1}^{L}}\Delta w_{i1} + O(||w_{i1}||_2^2)\\
&\approx f(w_{i1},x) + 0*\Delta w_{i1}\\
&=f(w_{i1},x) 
\end{split}
\end{align}

From Eq (\ref{eq:nodiff}) we conclude that a small perturbation of $w_{i1}$ will not be reflected in the output if $\alpha=0$. The same derivation also holds for $w_{i2}$ and $w_{i3}$. 

To address this problem, we propose our Maximum Active Neuron Cover (MANC) Sample Selection algorithm to select a small number of samples from a bag of generated \texttt{Sensitive-Samples} to avoid the inactive neurons. Our criterion is to minimize the number of neurons not being activated by any \texttt{Sensitive-Samples}, or equivalently, maximize the number of neurons being activated at least once by the selected samples. We call the resultant set of Sensitive-Samples with their correct model outputs, the \textbf{\textit{fingerprint}} of the DNN model.

We can abstract it to a maximum coverage problem \cite{ageev1999approximation}. As input, we are given a bag of generated \texttt{Sensitive-Samples} $B=\{S_1,...,S_N\}$ and $k$, the number of desired samples. Suppose each \texttt{Sensitive-Sample} $S_i$ activates a set of neurons $P_i$. The set $\{P_i\}$ may have elements (neurons) in common. We will select $k$ of these sets such that the maximum number of elements (neurons) are covered, i.e. the union of the selected sets has maximal size. 
We define the set of neurons being activated at least once by the $k$ samples as \textit{Active-Neuron Cover (ANC)}. It is the union of individually activated neurons $P_i$, i.e. $\bigcup_{i=1}^{k} P_{k}$. We would like to maximize the number of elements (neurons) in $ANC$, i.e. maximize $|\bigcup_{i=1}^{k} P_{k}|$.

\begin{figure}[h]
\centering
\includegraphics[width=0.5\textwidth]{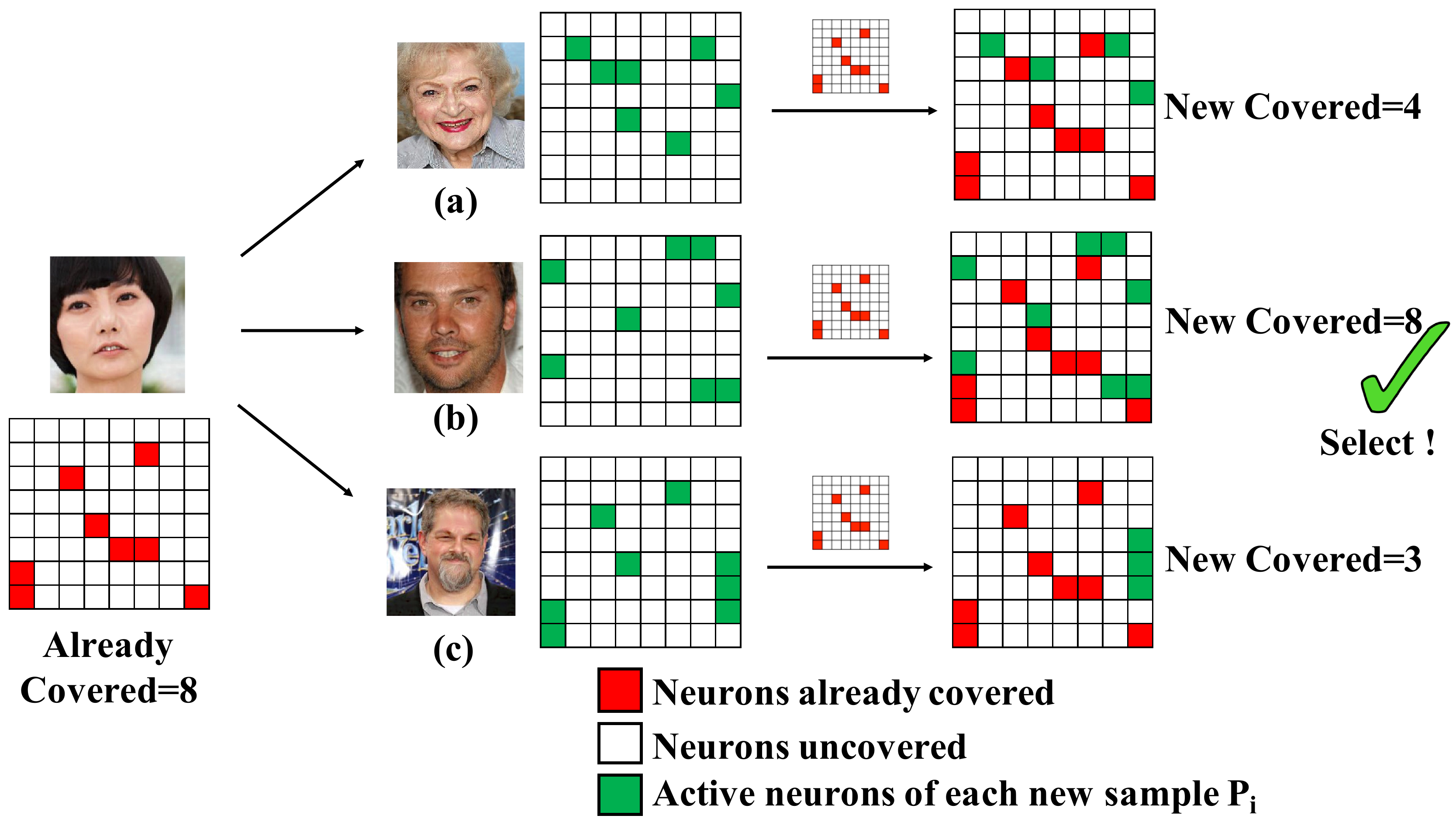}
\caption{Illustration of selecting one sample in Algorithm 2 (line 17-27). Suppose the set $Sample\_{List}$ initially contains one selected sample (young lady, left). We want to select the next sample from three candidates (a),(b) and (c). We compute the neurons (red) that have been activated by the samples already in $S$, i.e. Active-Neuron Cover, and the uncovered neurons (white). We also compute the neurons activated by each candidate ($P_i$). Candidate samples (a),(b) and (c) activate 4,8 and 3 uncovered neurons, respectively. Thus we add the candidate (b) to $Sample\_{List}$ and update the covered neurons.}
\label{fig:illustration-alg2}
\end{figure}

The generalized maximum coverage problem is known to be NP-hard \cite{feige1998threshold}, thus we use a greedy search to handle it. Intuitively, in each iteration $t$, we choose a set $P_t$ which contains the largest number of uncovered neurons. We show the pseudo-code of MANC algorithm in Algorithm 2. We show an illustration of one step in the MANC algorithm in Figure \ref{fig:illustration-alg2}. 

\setlength{\textfloatsep}{3pt}
\begin{algorithm}[h]
\caption{Maximum Active Neuron Cover(MANC) Sample Selection}\label{alg:MANC}
\begin{algorithmic}[1]
\State \textbf{Function} MANC(Neurons, B, k)
\\
\State /* Neurons: The neurons of interest */
\State /* B: The bag of samples from Algorithm 1 */
\State /* k: Number of desired samples */
\\
\State Uncovered = Neurons
\State Sample$\_$List = []
\\
\State /* Each sample $B[i]$ activates neurons $P_i$ */
\For{$(i=0; i<|B|; i++)$}
    \State $\alpha$ = Activation(Neurons, B[i])
    \State $P_i = \alpha > 0$
\EndFor \State \textbf{end for}
\\
\State /* Outer loop selects one sample each time */
\For{$(i=0; i<k; i++)$}
\\
    \State /* Inner loop among all samples to find the one that activates the largest number of uncovered neurons */
    \For{$(j=0; j<|B|; j++)$}
        \State NewCovered$_j$ = Uncovered $\bigcap$ $P_j$
        \State $N_j$ = $|\ NewCovered_j\ |$
    \EndFor \State \textbf{end for}
    \State l = $\argmax_j$ \ $N_j$
    \State Sample$\_$List.add($B[l]$)
    \State Uncovered = Uncovered - $P_l$
\EndFor \State \textbf{end for}
\State \textbf{return} Sample\_List
\end{algorithmic}
\end{algorithm}

Line 7 in Algorithm 2 initializes the uncovered neurons to all neurons of interest. Line 8 initializes the set of the selected sample to null. Line 12 computes the activations of neurons with corresponding input \texttt{Sensitive-Sample} $B[i]$. Line 13 determines the neurons that are activated by $B[i]$, i.e. $P_i$. Line 17 loops to select one sample in each iteration. Lines 20-25 determine which sample activates the largest number of uncovered neurons, and add it to the selected sample set. Line 26 updates the uncovered neurons.

\subsection{Model Output Specification}\label{sec:method:outputspec}

The form of the model output significantly affects the information that can be retrieved through black-box access. We consider three forms of $y$ as the outputs of a DNN for classification tasks:
\begin{packeditemize}
\item Case 1: Numerical probabilities of each class.
\item Case 2: Top-k (k$>$1) classification labels.
\item Case 3: Top-1 classification label.
\end{packeditemize} 

In general, the less information included in the output (from Case 1 (most) to Case 3 (least)), the harder to generate valid \texttt{Sensitive-Samples} and fingerprints. However, in our experiments, we observe that our proposed algorithm can detect an integrity breach for all known real attacks if only the top-1 label is provided (case 3) with high accuracy ($>$99.5\%, $<$10 samples). Our experiments also show that we need even fewer samples ($<$3 samples) if more information is provided (cases 1 and 2). We discuss these results in detail in Section \ref{sec:eval}.


\subsection{Adversarial Examples and Sensitive-Samples}


A similar and popular concept of our proposed \texttt{Sensitive-Samples} is adversarial examples \cite{SzZaSSu:13}: the adversary intentionally adds human unnoticeable permutation $\Delta x$ to the normal samples $x$, so the model gives a wrong prediction for this sample, i.e., $f_{\theta}(x+\Delta x) \neq f_{\theta}(x)$. The adversarial examples are commonly used in evasion attacks (Section \ref{sec:adversarysamples}).

In this paper, we introduce \texttt{Sensitive-Samples}, another type of transformed inputs. \texttt{Sensitive-Samples} also have human unnoticeable permutation from the normal samples, i.e., $z'=z+\Delta z$. Instead of making the model give wrong outputs, the outputs of the \texttt{Sensitive-Samples} change with the model parameters, i.e., $f_{\theta}(z') \neq f_{\theta+\Delta \theta}(z')$. Thus, unlike adversarial examples usually being used as an evasion attack strategy, \texttt{Sensitive-Samples} can be use as a powerful approach to defend against model integrity attack. Table \ref{table:compare} shows the comparisons between the \texttt{Sensitive-Samples} and adversarial examples.  



\begin{table}[ht]
\caption{Comparisons between \texttt{Sensitive-Samples} and adversarial examples.}
\label{table:compare}
\resizebox{\columnwidth}{!}{
\begin{tabular}{|c|c|c|}
\hline
           & Sensitive-Samples                                                   & Adversarial-Examples                                           \\ \hline
Similarity & \multicolumn{2}{c|}{Transformed inputs}                                                                                              \\ \hline
Purpose    & Defense                                                             & Attack                                                         \\ \hline
Settings   & \begin{tabular}[c]{@{}c@{}}Model parameters change\\ $f_{\theta}(z') \neq f_{\theta+\Delta \theta}(z')$ \end{tabular} & \begin{tabular}[c]{@{}c@{}}Input perturbation\\ $f_{\theta}(x+\Delta x) \neq f_{\theta}(x)$\end{tabular} \\ \hline
Generation & White-box                                                           & White/Black box                                                \\ \hline
Usage      & Black-box                                                           & Black-box                                                      \\ \hline
Optimization &  Maximize the sensitivity                   &   \multirow{2}{*}{Maximize the cost function${}^{\star}$}       \\
Goal         &  of output w.r.t model  parameters          &             \\ \hline
\end{tabular}
}
    \begin{tablenotes}
      \scriptsize
      \item ${}^{\star}$ There exists other approaches to generate adversarial examples.
    \end{tablenotes}
\end{table}

\section{Implementation}
\label{sec:implementation}

\subsection{Attack Coverage}

Our proposed method is generic and able to detect integrity breaches due to various attacks against DNN models. We evaluate this method on four categories of attacks, based on the attack goal and techniques of compromising models. These cover from very subtle model changes to significant changes.

\bheading{Neural network trojan attack.} The attack goal is to inject trojan into the model so it will mis-classify the samples containing a specific trigger \cite{Trojannn,GuDoGa:17}. To achieve this, given a pretrained DNN model, the adversary carefully selects one or some ``critical'' neurons which the outputs are highly dependent on. Then he reverses the training data from the model, and adds triggers to them. He modifies the weights on the path from the selected neurons to the last layer by retraining the model using the data with triggers.



\bheading{Targeted poisoning attack.} The attack goal is to force the model to mis-classify a target class. The adversary achieves this by poisoning the dataset with carefully-crafted malicious samples. We consider two types of such attacks: the first one is error-generic poisoning attack \cite{biggio2012poisoning,mei2015using,xiao2015feature}, in which the outputs of the compromised model for the target class can be arbitrary. The second one is error-specific poisoning attack \cite{munoz2017towards}: the adversary modifies the model to mis-classify the target class as a fixed class that he desires. 




\bheading{Model compression attack.} The attack goal is to compress the DNN model without affecting the model accuracy significantly, to save cloud storage for profit. There are different compression techniques to achieve this, e.g., pruning \cite{HaMaJ:16}, quantization \cite{GoLiYa:14}, low precision \cite{CoBeDa:14}. In our experiments, we detect the model compression by the low precision technique. Other compression techniques can be detected similarly. 




\bheading{Arbitrary weights modification.} We consider the most general scenario: the adversary can change the weights of any arbitrary neurons to arbitrary values. The goal is to investigate the capability of our approach in defending against general model integrity breaches.

\subsection{Datasets and Models}

We evaluate our solution on attacks against different tasks and DNN models to show its effectiveness and generality. For most of the integrity attacks, we use the same datasets and models as in the literature. In Table \ref{table:datamodel}, we list the model specifications, as well as the attack results. 

Original model accuracy denotes the classification accuracy of the original correct model. Attack goal shows the adversary's target of modifying the model. Attack technique indicates how the adversary modifies the model. Note that we do not make any specific assumption on attack technique, providing comprehensive protection against all types of model modification. Attack result shows the mis-classification rate for neural network trojan attack and accuracy degradation attack, and the compression rate for model compression attack.

\begin{table*}[t]
  \centering
  \caption{Datasets and models in our evaluation.}\label{table:datamodel}
  \resizebox{\linewidth}{!}{
  \begin{tabular}{|c|c|c|c|c|c|c|c|c|c|c|c|}
  \hline
    \multicolumn{2}{|c|}{} & \multirow{2}{*}{\textbf{Dataset}} & \multirow{2}{*}{\textbf{Task}} & \multirow{2}{*}{\textbf{Model}} & \multirow{2}{*}{\textbf{\# layers}} & \textbf{\# Conv} & \textbf{\# FC} & \textbf{Original} & \textbf{Attack} & \textbf{Attack} & \textbf{Attack} \\
    \multicolumn{2}{|c|}{} & & & & & \textbf{layers} & \textbf{layers} & \textbf{accuracy} & \textbf{goal} & \textbf{technique} & \textbf{result} \\ \hline
    \multicolumn{2}{|c|}{\textbf{Neural network}} & \multirow{2}{*}{VGG-Face} & Face & \multirow{2}{*}{VGG-16} & \multirow{2}{*}{16}& \multirow{2}{*}{13} & \multirow{2}{*}{3}& \multirow{2}{*}{74.8\%}& Misclassify inputs& Selective neural& \multirow{2}{*}{100\%}\\
    \multicolumn{2}{|c|}{\textbf{trojan attack}} & & recognition & & & & & & with triggers& retraining & \\ \hline
    & \multirow{2}{*}{\textbf{Error-generic}} & \multirow{2}{*}{GTSRB} & Traffic sign & \multirow{2}{*}{CNN} & \multirow{2}{*}{7}& \multirow{2}{*}{6} & \multirow{2}{*}{1}& \multirow{2}{*}{95.6\%}& Misclassify ``Stop''& Data& \multirow{2}{*}{98.6\%}\\
    \textbf{Targeted} &  & & recognition & & & & & & traffic sign& poisoning & \\ \cline{2-12}
    \textbf{poisoning} & {\multirow{2}{*}{\textbf{Error-specific}}} & \multirow{2}{*}{GTSRB} & Traffic sign & \multirow{2}{*}{CNN} & \multirow{2}{*}{7}& \multirow{2}{*}{6} & \multirow{2}{*}{1}& \multirow{2}{*}{95.6\%}& Misclassify ``Stop''& Data& \multirow{2}{*}{87.3\%}\\
    & & & recognition & & & & & & to ``Speed 100km''& poisoning & \\ \hline
    \multicolumn{2}{|c|}{\multirow{2}{*}{\textbf{Model compression}}} & \multirow{2}{*}{CIFAR-10} & Image & \multirow{2}{*}{CNN} & \multirow{2}{*}{7}& \multirow{2}{*}{6} & \multirow{2}{*}{1}& \multirow{2}{*}{87.6\%}& \multirow{2}{*}{Save storage}& Precision& 4x compression\\
    \multicolumn{2}{|c|}{} & & classification & & & & & & & reduction & 84.1\%\\ \hline
    \multicolumn{2}{|c|}{\textbf{Arbitrary weights}} & \multirow{2}{*}{AT\&T} & Face & \multirow{2}{*}{MLP} & \multirow{2}{*}{1}& \multirow{2}{*}{0} & \multirow{2}{*}{1}& \multirow{2}{*}{95.0\%}& General model& Arbitrary& \multirow{2}{*}{$\star$}\\
    \multicolumn{2}{|c|}{\textbf{modification}} & & recognition & & & & & & modification& modification & \\ \hline
  \end{tabular}
  }
        \begin{tablenotes}
      \scriptsize
      \item ${}^{\star}$ We evaluate it for general integrity, thus no attack success rate.
    \end{tablenotes}

\end{table*}

\subsection{Hyper-parameters and Configurations}

In our experiments, we set the learning rate to 1*$10^{-3}$. We choose ADAM as our optimizer. We set $itr\_Max$ to 1000. We consider all the weights in the last layer as parameters-of-interest $W$. This is because the last layer must be modified in all existing attacks, and the output is most sensitive to this layer. We also observe that setting a threshold $\epsilon$ to the Signal-to-Noise Ratio (SNR), i.e. $\frac{||v-v_0||}{||v_0||}$, is generally better than using l2-norm ($||v-v_0||$). This is because the SNR both considers the perturbation and relative data intensity.

We reproduce the above four categories of DNN integrity attacks, and implement our solution using Tensorflow 1.4.1. We run our experiments on a server with 1 Nvidia 1080Ti GPU, 2 Intel Xeon E5-2667 CPUs, 32MB cache, 64GB memory and 2TB hard-disk.

\section{Evaluation}
\label{sec:eval}


\subsection{Neural Network Trojan Attack}
\label{sec:eval:sec:trojan}
We evaluate the DNN trojan inserted through selective neurons modification \cite{Trojannn} on the VGG-Face model and dataset. VGG-Face is a standard deep learning model used in face recognition tasks.

\begin{figure}[ht]
\centering
\includegraphics[width=0.45\textwidth]{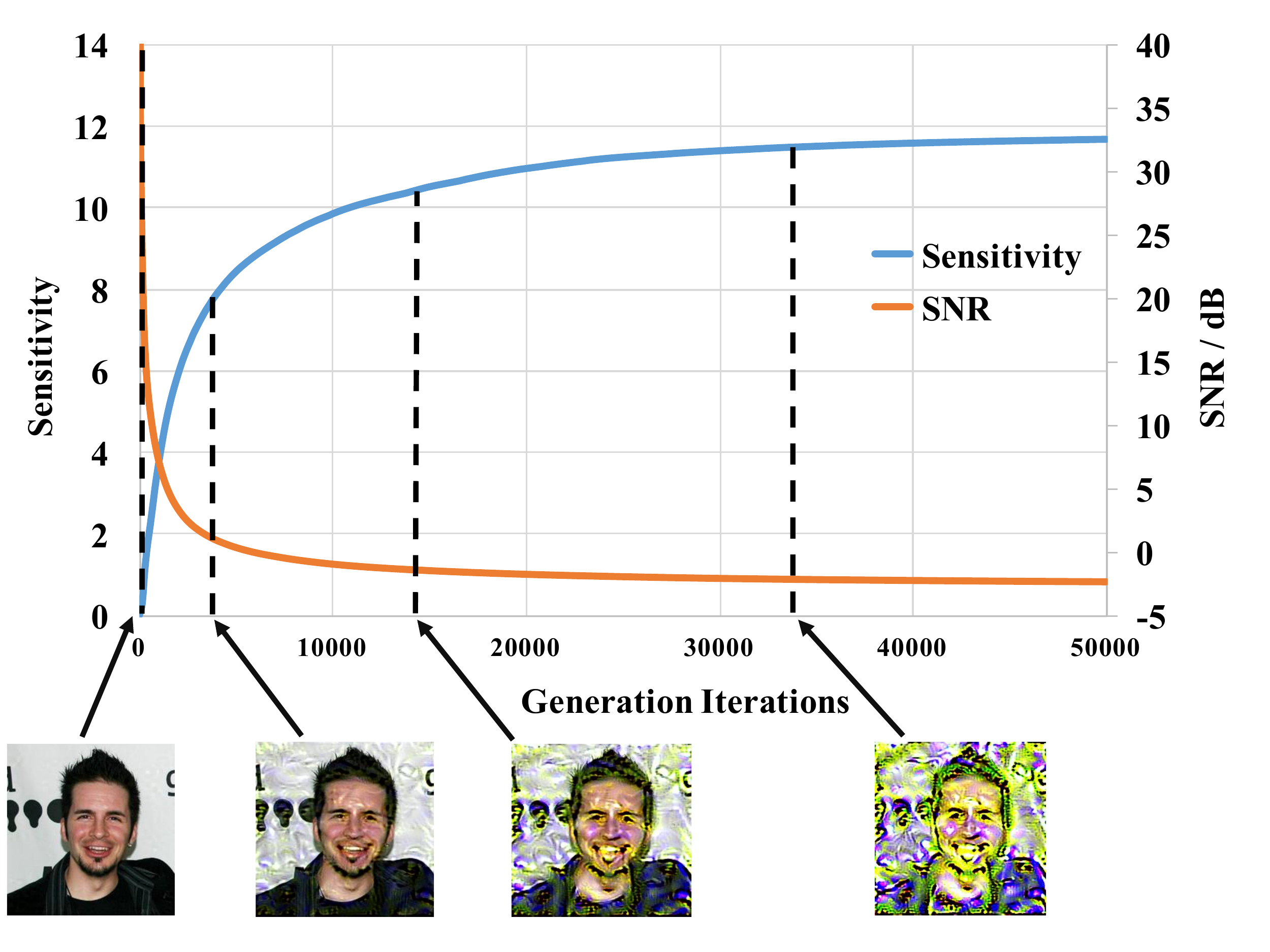}
\caption{Sensitivity and Signal-Noise Ratio (SNR) in the Sensitive-Sample generation process. As the optimization process goes on, the sensitivity of the generated image increases, while the SNR of the generated image decreases.}
\label{fig:vggface:mechanism}
\end{figure}

\bheading{Generation Mechanism.} Figure \ref{fig:vggface:mechanism} shows the trade-off between the sensitivity and SNR during the \texttt{Sensitive-Samples} generation process. The blue line represents the sensitivity, i.e. defined in Eq (\ref{eq:sensitivity}) as $||\frac{\partial f(W,x)}{\partial W}||_F^2$. The orange line represents the SNR. At the beginning of the optimization, the SNR is high, reflecting that the generated image is similar to the original input. However, the sensitivity is low, showing that the DNN output is not sensitive to the weight changes. It also indicates that directly using original images as fingerprints is not good. As the optimization goes on, the sensitivity increases significantly and finally converges to a high value. Meanwhile, artifacts are introduced in the sample generation, decreasing the similarity (SNR). We also show four intermediate images to show the trend of Sensitive-Sample generation in Figure \ref{fig:vggface:mechanism}. 

\begin{figure}[ht]
\centering
\includegraphics[width=0.4\textwidth]{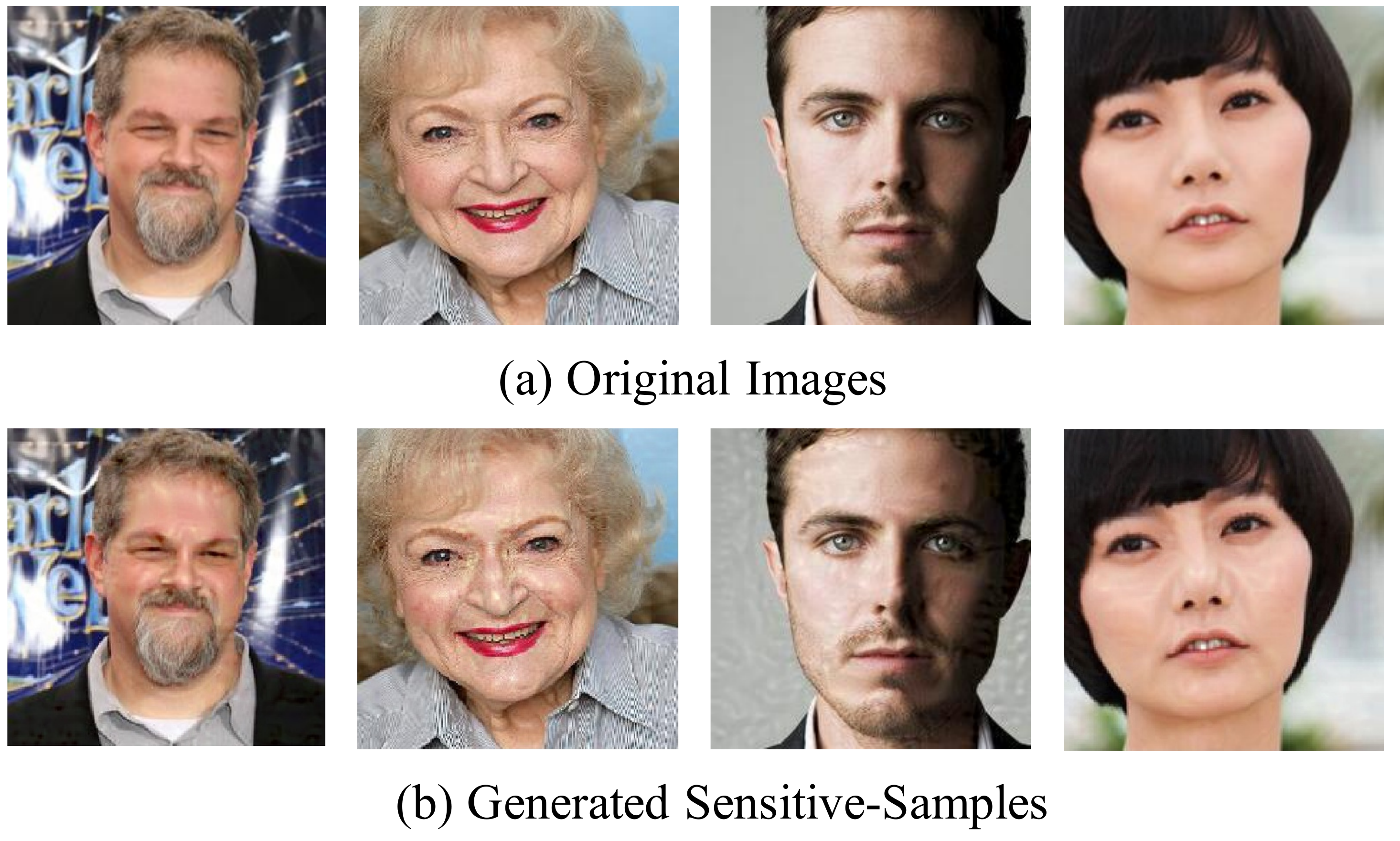}
\caption{Original and generated Sensitive-Sample images for integrity checking (Face Recognition)}\label{fig:vggface:gen-examples}
\end{figure}

In Figure \ref{fig:vggface:gen-examples}, we show representative examples of the \texttt{sensitive-samples}.  The generated images are very similar to the original inputs, with mild artifacts on the faces. Therefore, the attacker can hardly determine whether it is a natural image or a testing image for integrity checking.

\begin{figure}[ht]
\centering
\includegraphics[width=0.5\textwidth]{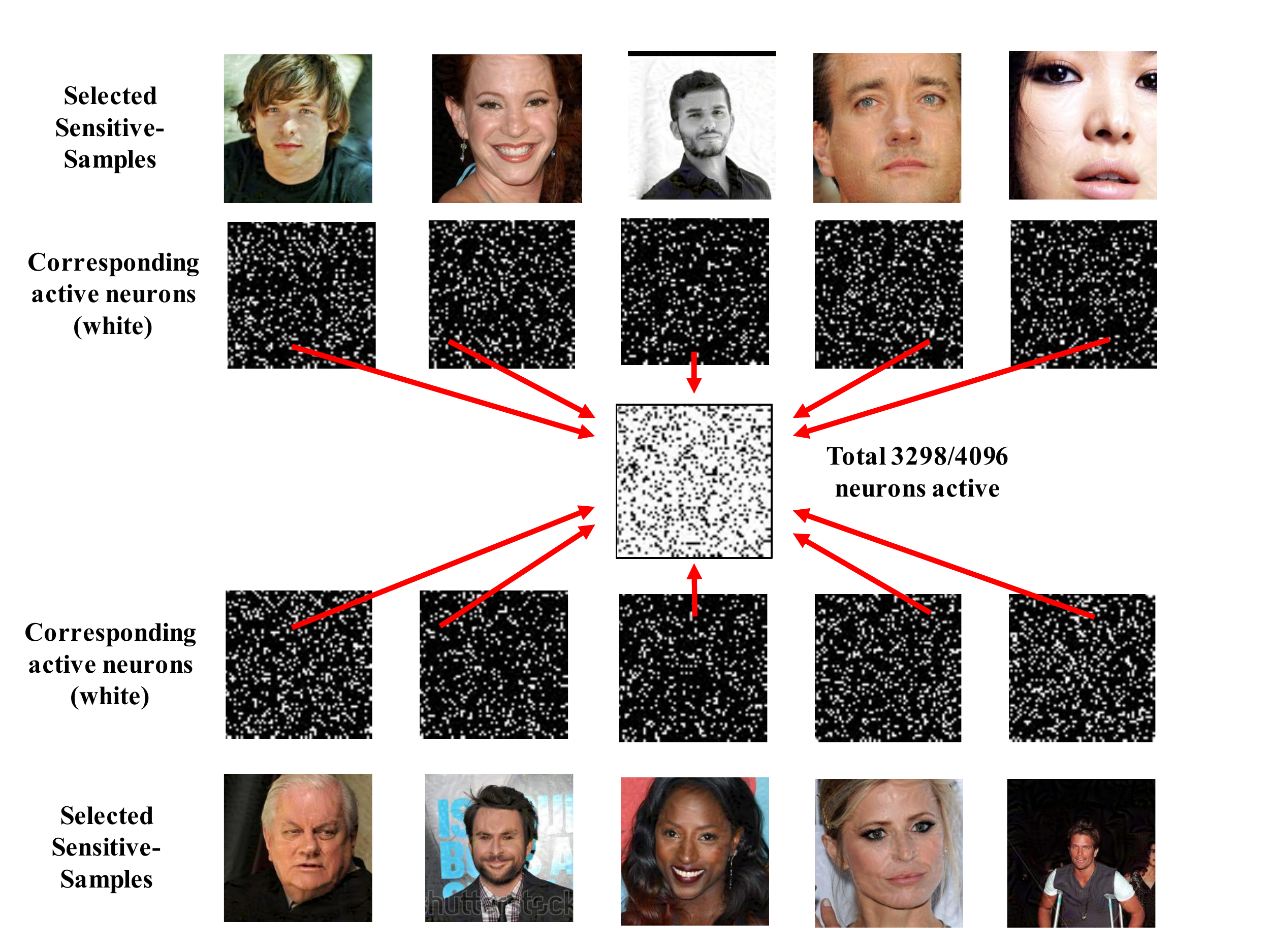}
\caption{An example of Active Neuron Cover (ANC)}
\label{fig:vggface:activeneurons}
\end{figure}

\bheading{Active-Neuron Cover.} We show a real example of the Active Neuron Cover (ANC) obtained from MANC in protecting the VGG-Face model in Figure \ref{fig:vggface:activeneurons}. We represent ten selected Sensitive-Samples in the top and bottom rows. The corresponding individual active-neurons $P_i$ are marked white in the two intermediate rows. The middle row shows the final ANC of the selected Sensitive-Samples. In this example, 3296/4096=80.5\% neurons are activated.

\begin{figure}[ht]
\centering
\includegraphics[width=0.4\textwidth]{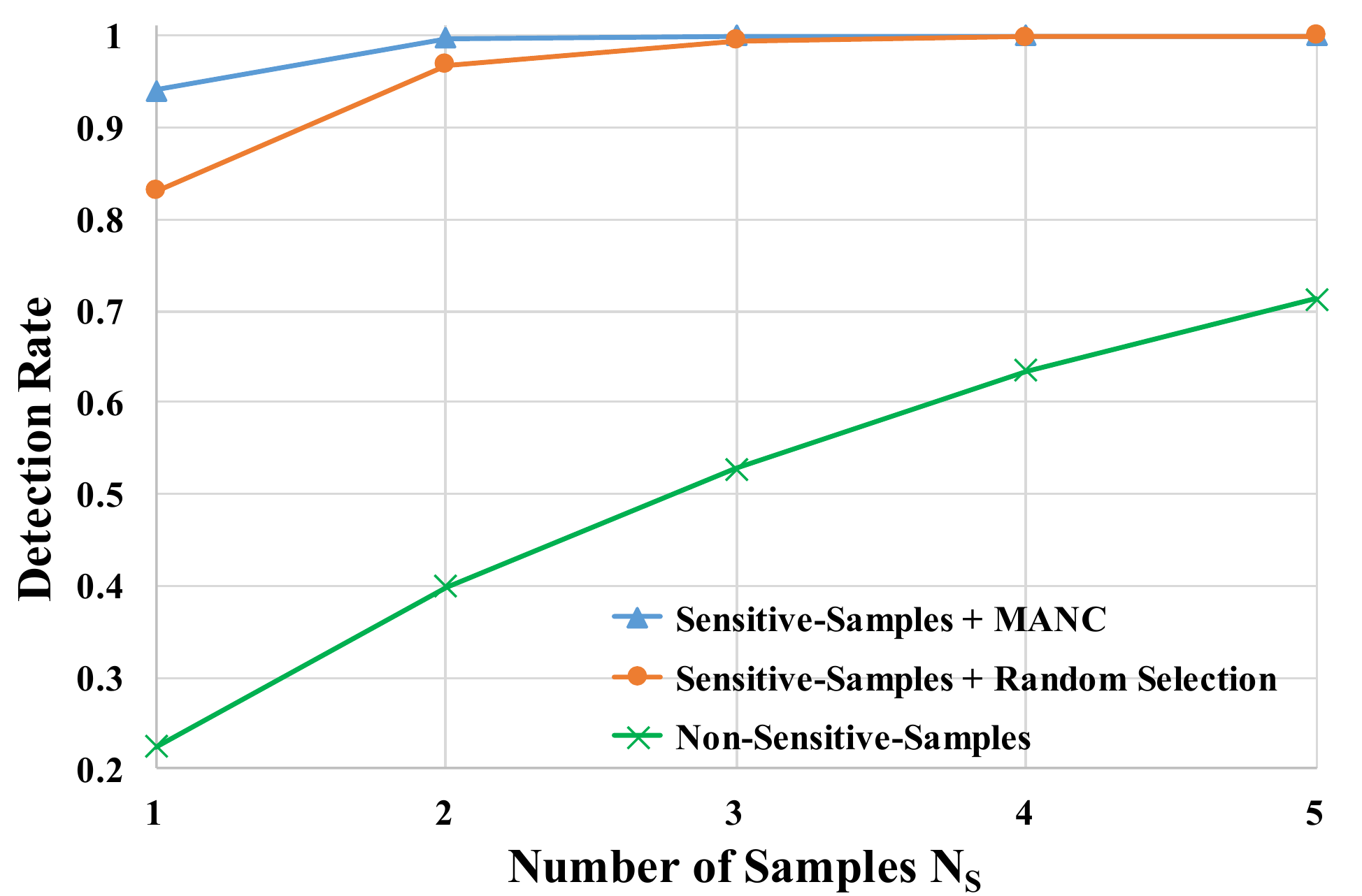}
\caption{Detection rate corresponds to the number of samples $N_S$ on VGG-Face. High detection rate ($>$99.8\%) is achieved with only 2 MANC selected Sensitive-Samples.}
\label{fig:vggface:detectionrate}
\end{figure}

\bheading{Detection Accuracy.} We define a successful detection as ``given $N_S$ sensitive samples, there is at least one sample, whose top-1 label predicted by the compromised model is different from the top-1 label predicted by the correct model''. Note that this is the most challenging case discussed in Section \ref{sec:method:outputspec}. We show the detection rate of (1) Sensitive-Samples with MANC sample selection (blue), (2) Sensitive-Samples with random selection (orange) and (3) Non-Sensitive Samples (green) in Figure \ref{fig:vggface:detectionrate}. In case (1) and (2), we first generate a bag of 100 sensitive-samples and select $N_S$ of them using MANC and random selection, respectively. In case (3), we randomly select $N_S$ images from the original validation set. We repeat the experiment 10,000 times and report the average detection rate. We observe that Sensitive-Samples + MANC always achieves a higher detection rate than Sensitive-Samples + random selection. Sensitive-Samples based approaches are always much better than non-sensitive samples, regardless of $N_S$. Sensitive-Samples + MANC reaches 99.8\% and 100\% detection rate at $N_S=2,3$ respectively. It indicates that we only need very few \texttt{sensitive-samples} ($<$4) to detect the model integrity breach.

\bheading{Output Specification.} We list the detection rates corresponding to different output specifications (columns) and $N_S$ (rows) in Table \ref{table:vggface:topk}. ``top-$k$'' means the model outputs the $k$ top labels. ``p-dec-$n$'' means the model outputs probabilities in addition to labels, with $n$ numbers after the decimal point. For example, ``Top-1-p-dec-2'' means the model outputs top-1 probability with the precision of 2 numbers after the decimal point. Comparing the first 3 columns, a larger $k$ provides higher detection rates because more information is embedded in the output. Comparing columns 1 and 4, the additional information of class probabilities helps increase the detection rate. Comparing the last two columns, the higher precision of the per-class probability also increases the detection rate. 

\begin{table}[ht]
\centering
\caption{Detection rates (\%) w.r.t to the output specifications. Large $k$, probability information and high precision of the probabilities can increase the trojan detection rate.}
\label{table:vggface:topk}
\resizebox{\columnwidth}{!}{%
\begin{tabular}{|c|c|c|c|c|c|c|}
\hline
$\#$ of samples  & top-1 & top-3 & top-5 & top-1-p-dec2 & p-dec-1 & p-dec-2 \\ \hline
1 & 94.1 & 100.0 & 100.0 & 99.6        & 99.8   & 100.0  \\ \hline
2 & 99.8 & 100.0 & 100.0 & 100.0        & 100.0   & 100.0  \\ \hline
3 & 100.0 & 100.0 & 100.0 & 100.0        & 100.0   & 100.0  \\ \hline
4 & 100.0 & 100.0 & 100.0 & 100.0        & 100.0   & 100.0  \\ \hline
5 & 100.0 & 100.0 & 100.0 & 100.0        & 100.0   & 100.0  \\ \hline
\end{tabular}
}
\end{table}

\bheading{False Positives.} Our proposed Sensitive-Samples defense leverages the determinacy of DNN model inference, therefore no false positive is raised. It is also true for all the models and datasets we evaluate below. We claim this is another advantage of our solution.

\subsection{Targeted Poisoning Attack}
\label{sec:eval:sec:traffic}

We evaluate our proposed method on detecting model changes through targeted poisoning attacks. 
We consider two attack scenarios. (1) Error-generic attack: the attacker's goal is to maximize the degradation of the prediction accuracy of the target class. The attacker does not care about the resultant output, as long as it is incorrect. (2) Error-generic attack: the attacker's goal is to mis-classify the samples from the target class to a specific class that he desires. 



\eheading{1) Error-generic attack.}

We evaluate the targeted error-generic attack on a security-critical application, the traffic sign recognition system used in autonomous-driving cars. The target model is a 7-layer CNN model trained over the German Traffic Sign Recognition Benchmark (GTSRB) database, which can recognize more than 40 classes of traffic signs. The attacker slightly modifies the model through data poisoning, in order to degrade the classification accuracy of ``STOP'' sign, while the recognition accuracy of all other traffic signs remains the same. 

\begin{figure}[ht]
    \centering
    \begin{subfigure}[ht]{\linewidth}
        \centering
        \includegraphics[width=0.1\linewidth]{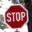}
        \includegraphics[width=0.1\linewidth]{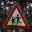}
        \includegraphics[width=0.1\linewidth]{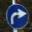}
        \includegraphics[width=0.1\linewidth]{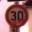}
        \includegraphics[width=0.1\linewidth]{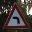}
        \includegraphics[width=0.1\linewidth]{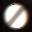}
        \includegraphics[width=0.1\linewidth]{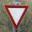}
        \includegraphics[width=0.1\linewidth]{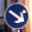}
        \caption{Original Images}
        \label{fig:traffic:orig}
    \end{subfigure}
    \begin{subfigure}[ht]{\linewidth}            
        \centering         
        \includegraphics[width=0.1\linewidth]{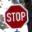}
        \includegraphics[width=0.1\linewidth]{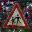}
        \includegraphics[width=0.1\linewidth]{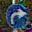}
        \includegraphics[width=0.1\linewidth]{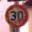}
        \includegraphics[width=0.1\linewidth]{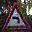}
        \includegraphics[width=0.1\linewidth]{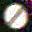}
        \includegraphics[width=0.1\linewidth]{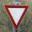}
        \includegraphics[width=0.1\linewidth]{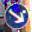}
        \caption{Generated Sensitive-Samples}
        \label{fig:traffic:gen}
    \end{subfigure}
    \caption{Original and generated Sensitive-Samples for integrity protection (GTSRB Traffic Sign)}\label{fig:traffic:gen-examples}
\end{figure}

\bheading{Generation Mechanism.}
We use Sensitive-Sample fingerprinting to verify if the traffic sign recognition is unchanged. 
Figure \ref{fig:traffic:gen-examples} shows some examples on the GTSRB dataset. Original images are presented in Figure \ref{fig:traffic:orig}. The corresponding generated images are presented in Figure \ref{fig:traffic:gen}.

\begin{figure}[ht]
\centering
\includegraphics[width=0.45\textwidth]{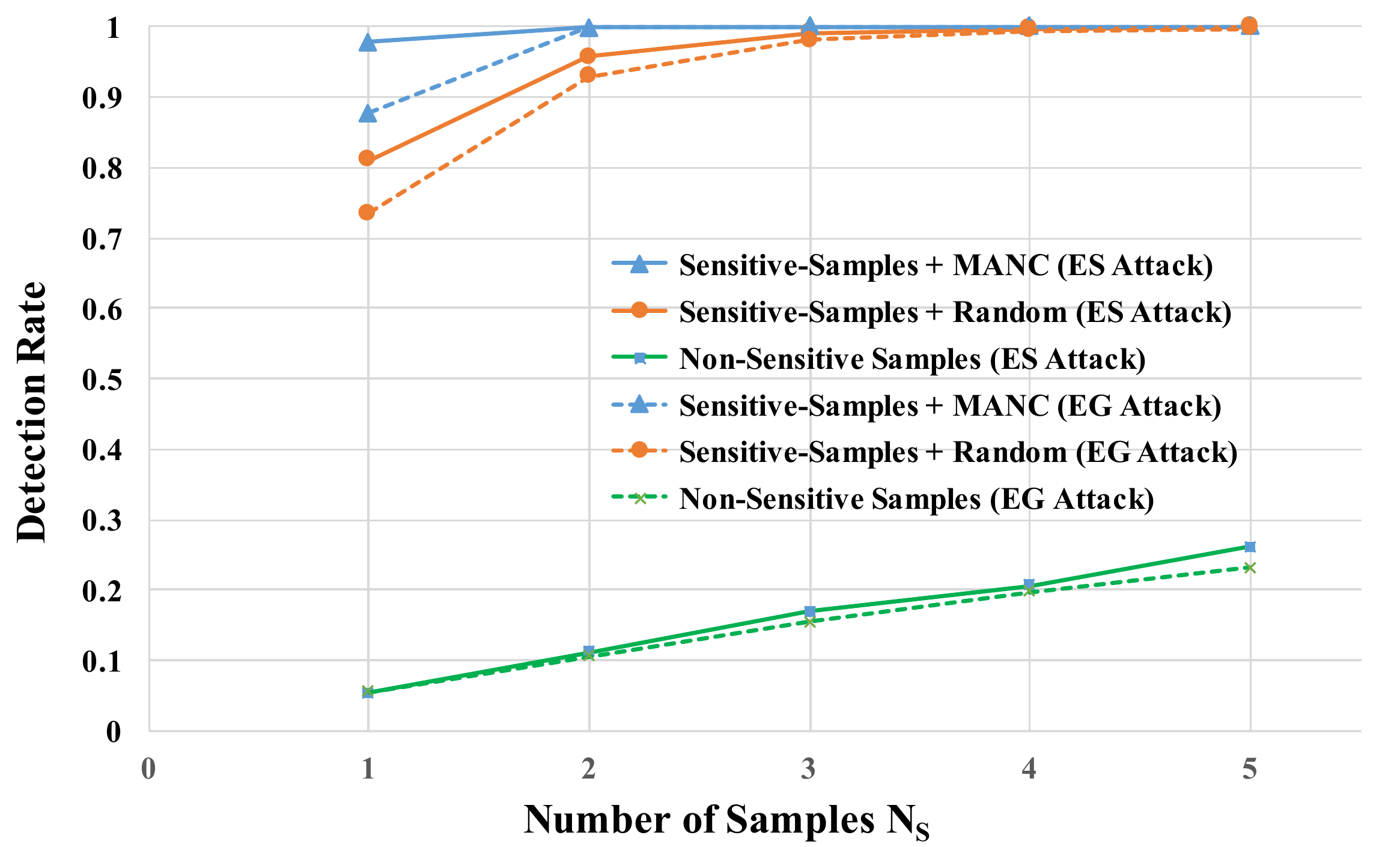}
\caption{Detection rate w.r.t the number of samples on the GTSRB dataset. Dotted lines represent error-generic attack. Solid lines represent error-specific attack. Sensitive-Samples+MANC (blue) achieves significantly higher detection rate than non-sensitive sample (green).}
\label{fig:traffic:detectionrate}
\end{figure}

\bheading{Detection Accuracy.} We show the detection rate of Sensitive-Samples + MANC (dotted blue), Sensitive-Samples + random selection (dotted orange) and Non-Sensitive Samples (dotted green) in Figure \ref{fig:traffic:detectionrate}. The detection rate increases with more samples. MANC improves the detection rate by 15\% when $N_S=1$, and 7\% when $N_S=2$. When $N_S=2$, Sensitive-Samples + MANC achieves the detection rate 100\%. Sensitive-Samples + random selection achieves 92.9\% and Non-Sensitive Samples only gets 10.6\%. When $N_S>5$, both Sensitive-Samples + MANC and Sensitive-Samples + random selection achieve 100\% accuracy. However, non-sensitive samples detect only 41.6\% model changes. This comparison shows the effectiveness of our approach. 


\bheading{Output Specification.} We show the detection rate corresponding to the output specifications in Table \ref{table:traffic:topk}. Comparing columns 1,2 and 3, we conclude that the larger $k$ provides more information and helps achieve higher detection rate. Probabilities provide extra information and also increase the detection rate, by comparing columns 1 and 5. However, this improvement (1.1\%) is not as significant as in the previous VGG-Face recognition dataset (5\%), because most of the probabilities are small and all rounded to 0.0, thus removing (rather than adding) information, compared to columns 2 and 3. For the same reason, outputting the probabilities of all classes with 2 digits (p-dec-2) after the decimal point does not improve the detection rate, compared to 1 digits (p-dec-1).

\begin{table}[ht]
\centering
\caption{Detection rates (\%) w.r.t to the output forms on GTSRB Traffic Sign dataset. In an error-generic attack, the ``STOP'' signs are mis-classified as a random wrong class. Large k, probability information and high precision of the probabilities help with the model modification detection.}
\label{table:traffic:topk}
\resizebox{\columnwidth}{!}{%
\begin{tabular}{|c|c|c|c|c|c|c|}
\hline
$\#$ of samples  & top-1 & top-3 & top-5 & top-1-p-dec2 & p-dec-1 & p-dec-2 \\ \hline
1 & 87.68 & 100.0 & 100.0 & 88.76        & 88.76   & 88.76  \\ \hline
2 & 99.97 & 100.0 & 100.0 & 99.87        & 99.98   & 100.0  \\ \hline
3 & 99.99 & 100.0 & 100.0 & 100.0        & 99.99   & 100.0  \\ \hline
5 & 100.0 & 100.0 & 100.0 & 100.0        & 100.0   & 100.0  \\ \hline
8 & 100.0 & 100.0 & 100.0 & 100.0        & 100.0   & 100.0  \\ \hline
10 & 100.0 & 100.0 & 100.0 & 100.0        & 100.0   & 100.0  \\ \hline
\end{tabular}
}
\end{table}

\begin{table}[ht]
\centering
\caption{Detection rates (\%) w.r.t to the output forms on GTSRB Traffic Sign dataset. In an error-specific attack, the ``STOP'' signs are mis-classified as ``Speed limit 100 km/h''. Large k, probability information and high precision of the probabilities help with the model modification detection.}
\label{table:traffic-error-specific:topk}
\resizebox{\columnwidth}{!}{%
\begin{tabular}{|c|c|c|c|c|c|c|}
\hline
$\#$ of samples  & top-1 & top-3 & top-5 & top-1-p-dec2 & p-dec-1 & p-dec-2 \\ \hline
1 & 97.90 & 100.0 & 100.0 & 100.0        & 100.0   & 100.0  \\ \hline
2 & 99.97 & 100.0 & 100.0 & 100.0        & 100.0   & 100.0  \\ \hline
3 & 100.0 & 100.0 & 100.0 & 100.0        & 100.0   & 100.0  \\ \hline
5 & 100.0 & 100.0 & 100.0 & 100.0        & 100.0   & 100.0  \\ \hline
8 & 100.0 & 100.0 & 100.0 & 100.0        & 100.0   & 100.0  \\ \hline
10 & 100.0 & 100.0 & 100.0 & 100.0        & 100.0   & 100.0  \\ \hline
\end{tabular}
}
\end{table}

\eheading{2) Error-specific Attack}

We also use the same CNN model and GTSRB database for evaluating the error-specific attack. The attacker adopts the data poisoning technique to modify the model, so the compromised model will mis-classify the ``STOP'' sign as ``Speed limit 100km/h'', while giving correct results for other traffic signs. 

\bheading{Detection Accuracy.} We show the detection rate of Sensitive-Samples + MANC (solid blue), Sensitive-Samples + random selection (solid orange) and Non-Sensitive Samples (solid green) in Figure \ref{fig:traffic:detectionrate}. We observe that the detection rates are higher than the error-generic attack, regardless of approaches and $N_S$. We can detect 97.9\% model modification with a single \texttt{Sensitive-Sample}. On the contrary, a non-sensitive sample only detects $<$ 5.0\% model changes. 

\bheading{Output Specification.} We show the detection rate corresponding to the output specifications in Table \ref{table:traffic-error-specific:topk}. Comparing columns 1,2 and 3, we conclude that the larger $k$ provides more information and helps achieve higher detection rate. We observe that the detection rate of error-specific attack is higher than error-generic attack using the same output specification. We also observe that, a single \texttt{Sensitive-Sample} can detect error-specific attack with 100\% accuracy if extra information, e.g. top-1 probabilities, top-k (k$>1$) classes and probabilities of all classes, is provided.

\subsection{Model Compression Attack}
\label{sec:eval:sec:compression}

Next we evaluate the detection of model integrity breach due to model compression. The target model is a CNN trained on CIFAR-10 dataset. To compress the model, the attacker stores the model with low precision, i.e., the parameters are converted from 32 bits to 8 bits. The model size is thus reduced by 4x, while the accuracy is reduced only by $3.6\%$.

\begin{figure}[ht]
    \centering
    \begin{subfigure}[ht]{\linewidth}   
        \centering         
        \includegraphics[width=0.15\linewidth]{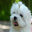}
        \includegraphics[width=0.15\linewidth]{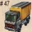}
        \includegraphics[width=0.15\linewidth]{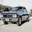}
        \includegraphics[width=0.15\linewidth]{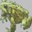}
        \includegraphics[width=0.15\linewidth]{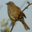}
        \includegraphics[width=0.15\linewidth]{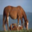}
        \caption{Original Images}
    \end{subfigure}

    \begin{subfigure}[ht]{\linewidth}            
        \centering         
        \includegraphics[width=0.15\linewidth]{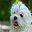}
        \includegraphics[width=0.15\linewidth]{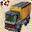}
        \includegraphics[width=0.15\linewidth]{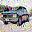}
        \includegraphics[width=0.15\linewidth]{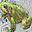}
        \includegraphics[width=0.15\linewidth]{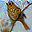}
        \includegraphics[width=0.15\linewidth]{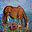}
        \caption{Generated Sensitive-Samples}
    \end{subfigure}
    \caption{Original and generated Sensitive-Samples for integrity protection (CIFAR)}\label{fig:compression:gen-examples}
\end{figure}

\bheading{Generation Mechanism.}
In Figure \ref{fig:compression:gen-examples}, we show the generated \texttt{sensitive-samples}. We observe that the generated samples look similar to the original ones. 



\begin{figure}[ht]
\centering
\includegraphics[width=0.40\textwidth]{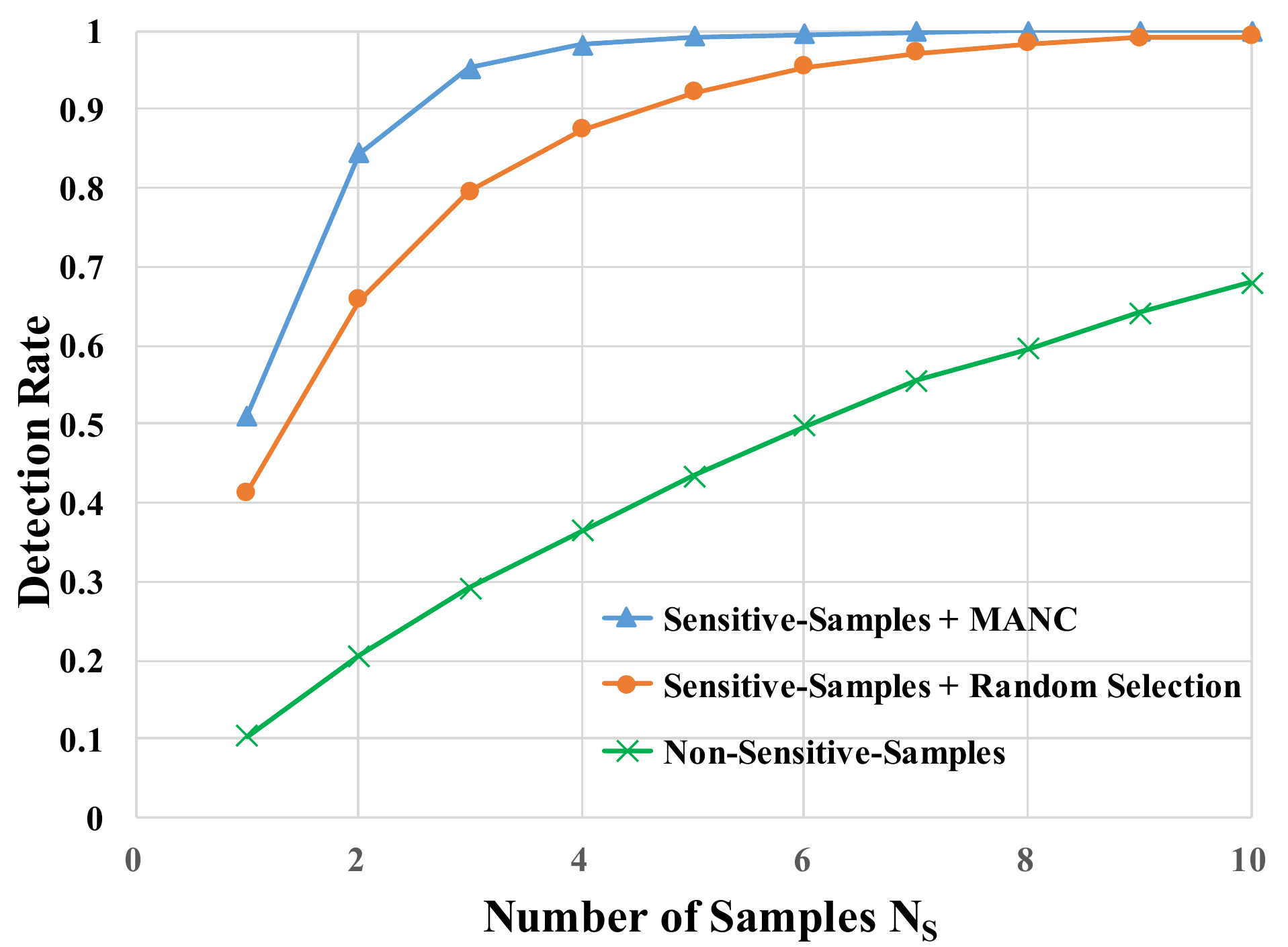}
\caption{Detection rate of model compression corresponds to the number of samples on CIFAR-10 dataset.}
\label{fig:compression:detectionrate}
\end{figure}

\bheading{Detection Accuracy.} 
We show the model compression detection rate of Sensitive-Samples + MANC (blue), Sensitive-Samples + random selection (orange) and Non-Sensitive Samples in Figure \ref{fig:compression:detectionrate}. The detection rate achieves 99.5\% when $N_S=6$ and ultimately $>$99.9\% when $N_S=10$. We observe that, our proposed Sensitive-Samples + MANC approach always outperforms other approaches, regardless of $N_S$.

\bheading{Output Specification.} Table \ref{table:compression:topk} shows the effect of output specification on the CIFAR-10 dataset. We observe that the detection rates of all output specifications increase as the number of Sensitive-Samples increases. The detection rate of model compression with probability outputs (last 3 columns in Table \ref{table:compression:topk}) are lower than the corresponding results in the previous model modifications (last 3 columns in Table \ref{table:vggface:topk}). It is because the numerical precision of decimal numbers are reduced when the model is compressed, resulting in two probabilities with small difference become indistinguishable. We also find that the detection rate when the model outputs top-1 probability with 2 decimal digits is lower than outputting top-1 index, when $N_S \leq 5$. This is because the model is compressed such that most of the top-1 probabilities are rounded to 100\%. One Sensitive-Sample achieves 100\% detection rate if the model outputs top-3 or top-5 indexes. In all other cases, 6 samples are enough to achieve $>$99.5\% detection rate.

\begin{table}[ht]
\centering
\caption{Detection rates (\%) w.r.t to the output forms on CIFAR-10 dataset. Large k, probability information and high precision of the probabilities help with the trojan detection.}
\label{table:compression:topk}
\resizebox{\columnwidth}{!}{%
\begin{tabular}{|c|c|c|c|c|c|c|}
\hline
$\#$ of samples     & top-1     & top-3        & top-5     & top-1-p-dec2 & p-dec-1    & p-dec-2 \\ \hline
1                   & 50.29     & 100.0        & 100.0     & 49.79        & 54.07      & 71.21  \\ \hline
2                   & 84.68     & 100.0        & 100.0     & 69.92        & 89.13      & 94.45  \\ \hline
3                   & 95.23     & 100.0        & 100.0     & 87.94        & 97.71      & 99.25  \\ \hline
5                   & 99.14     & 100.0        & 100.0     & 98.92        & 99.78      & 99.99  \\ \hline
8                   & 99.87     & 100.0        & 100.0     & 99.98        & 100.0      & 100.0  \\ \hline
10                  & 99.96     & 100.0        & 100.0     & 100.0        & 100.0      & 100.0  \\ \hline
\end{tabular}
}
\end{table}

\subsection{Arbitrary Weights Modification}
\label{sec:eval:sec:arbit}

In addition to specific model integrity attacks, we discuss the feasibility of our proposed defense against any arbitrary DNN integrity breaches. The adversary can arbitrarily modify a subset of the weights. We would like to investigate how the detection rate changes w.r.t the ratio of changed weights.

To simulate this attack, we select a ratio $r$ (0.1\%-80\%) and randomly modify $r$ weights, by adding Gaussian noise with zero mean and unit standard deviation. Note that most model integrity attacks need to modify all weights. Other attack techniques, e.g., DNN trojan via neuron selection, at least have to modify the fully connected layers, which changes more weights than we evaluate in this section.

To eliminate the effect of the unbalanced importance of weights on different layers, we choose a single-layer perceptron as our evaluated model. The weights on the same layer contribute nearly the same in a model. We evaluate the model on the AT\&T face recognition dataset. There are 40 distinct classes in the dataset. 

\begin{figure}[ht]
\centering
\includegraphics[width=0.45\textwidth]{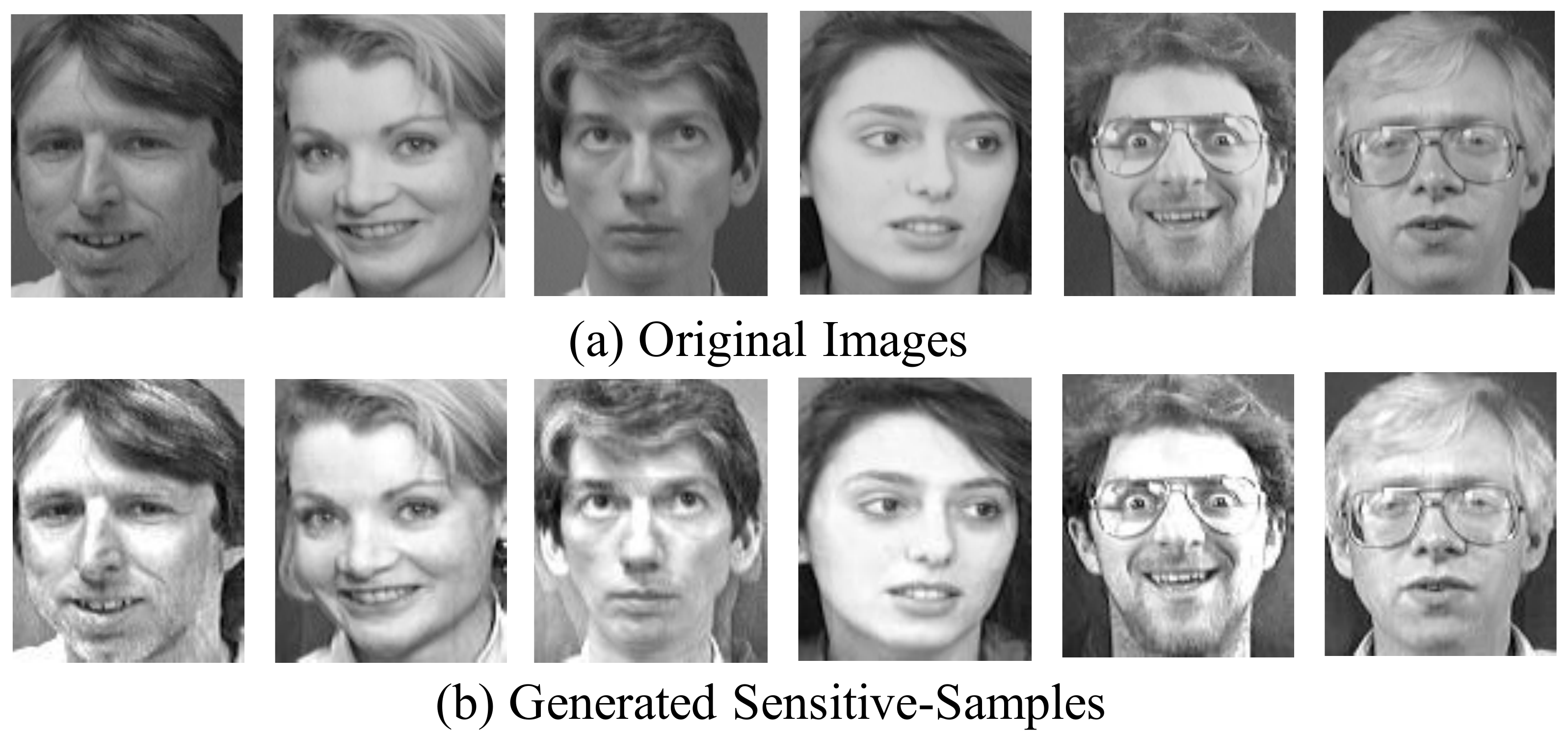}
\caption{Generated Sensitive-Samples for AT\&T dataset.}
\label{fig:arbit:gen-examples}
\end{figure}

\bheading{Generation Mechanism.} We show the generated Sensitive-Samples in Figure \ref{fig:arbit:gen-examples}. Few artifacts are introduced in the samples. However, the brightness of the generated samples is enhanced. We find this is because many pixels are truncated to the maximum intensity, i.e. 255, in the constraint set projection step in Algorithm 1. On the other hand, these generated images are still very similar to the natural ones.

\begin{figure}[ht]
\centering
\includegraphics[width=0.4\textwidth]{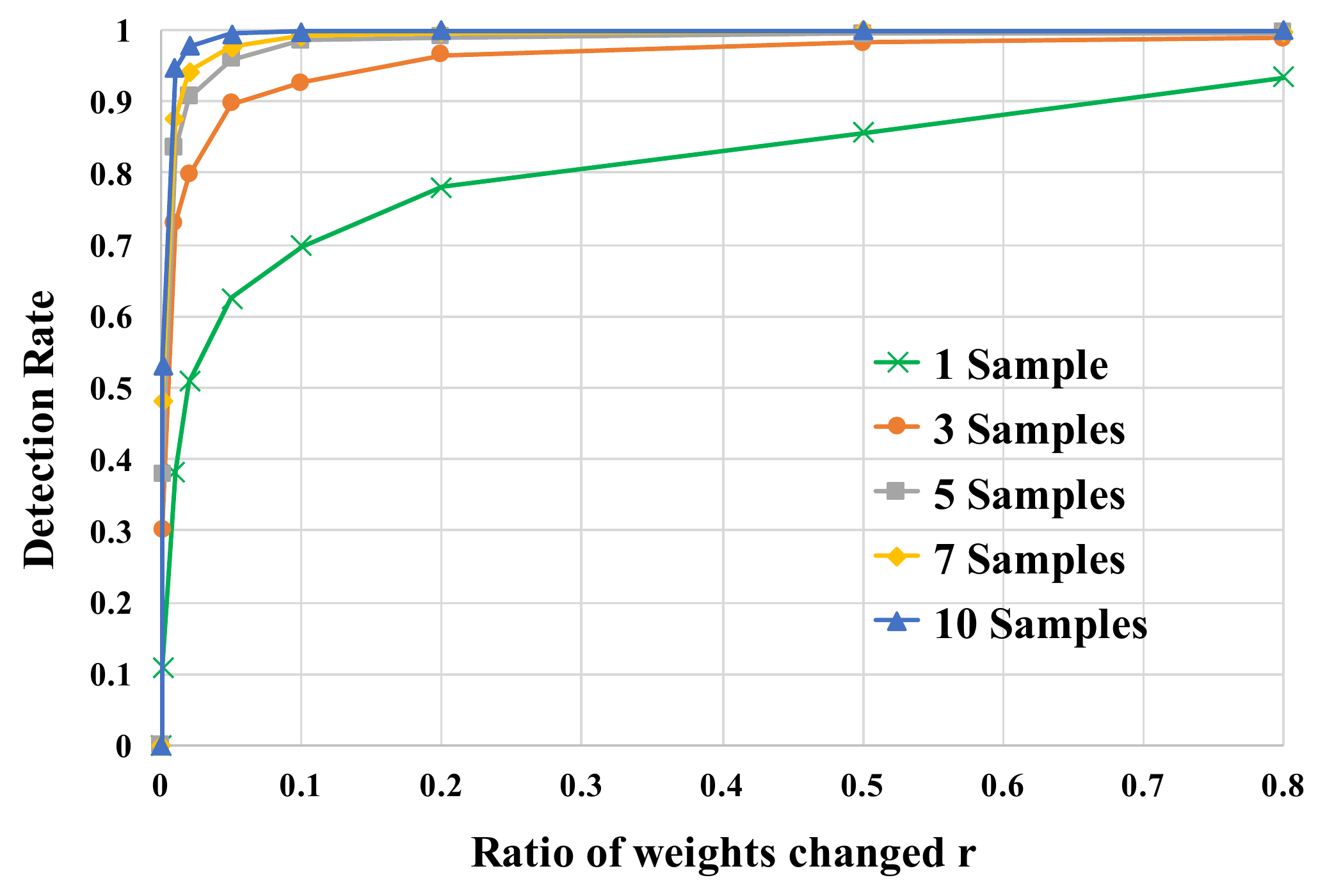}
\caption{Detection rate w.r.t the ratio of weights changed.}
\label{fig:cifar:topk}
\end{figure}

\bheading{Detection Accuracy.} The resultant detection rates versus $r$ are presented in Figure \ref{fig:cifar:topk}. Clearly, the detection rate increases as $r$ increases. The detection rate also increases as more Sensitive-Samples are leveraged. When $r$=1\%, i.e one order smaller than the number of changed weights in real attacks, our proposed method detects $>$95\% of the integrity breaches.

\section{Discussion}
\label{sec:discussion}

\subsection{Alternative Image Transforms}


In Section \ref{sec:eval}, we compare the effectiveness of our proposed \texttt{Sensitive-Samples} with randomly selected normal samples. Here we discuss some alternative transforms for DNN integrity verification. We apply some transforms on the input samples to generate fingerprints of the DNN models. We consider \ding{172} adding random noise, \ding{173} rotating the image and \ding{174} distorting the image. 

\begin{figure}[h]
    \centering
    \includegraphics[width=0.18\linewidth]{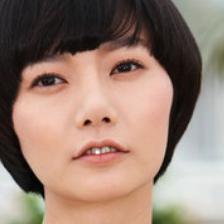}
    \includegraphics[width=0.18\linewidth]{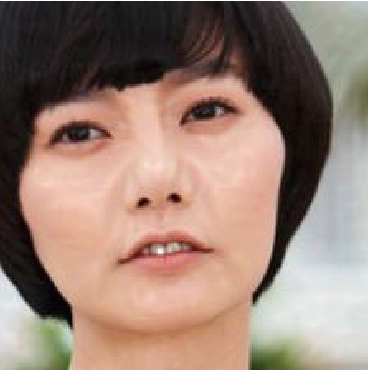}
    \includegraphics[width=0.18\linewidth]{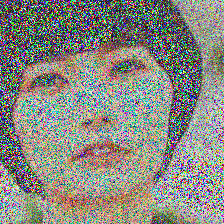}
    \includegraphics[width=0.18\linewidth]{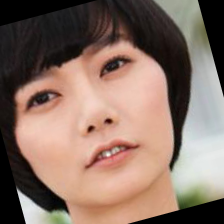}
    \includegraphics[width=0.18\linewidth]{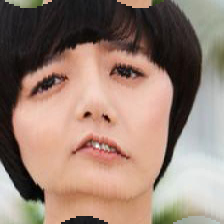}
    \caption{Benchmarks from left to right: The original image, Sensitive-Sample (Proposed), noised image, rotated image and distorted image. }\label{fig:benchmark:imgs}
\end{figure}

We use VGG-Face dataset as an example. Figure \ref{fig:benchmark:imgs} shows the original sample, \texttt{Sensitive-Sample}, and samples with noise, rotation and distortion operations. We observe that the \texttt{Sensitive-Sample} is the most similar one to the original image, thus imperceptible to the adversaries. 

\begin{figure}[h]
\centering
\includegraphics[width=0.45\textwidth]{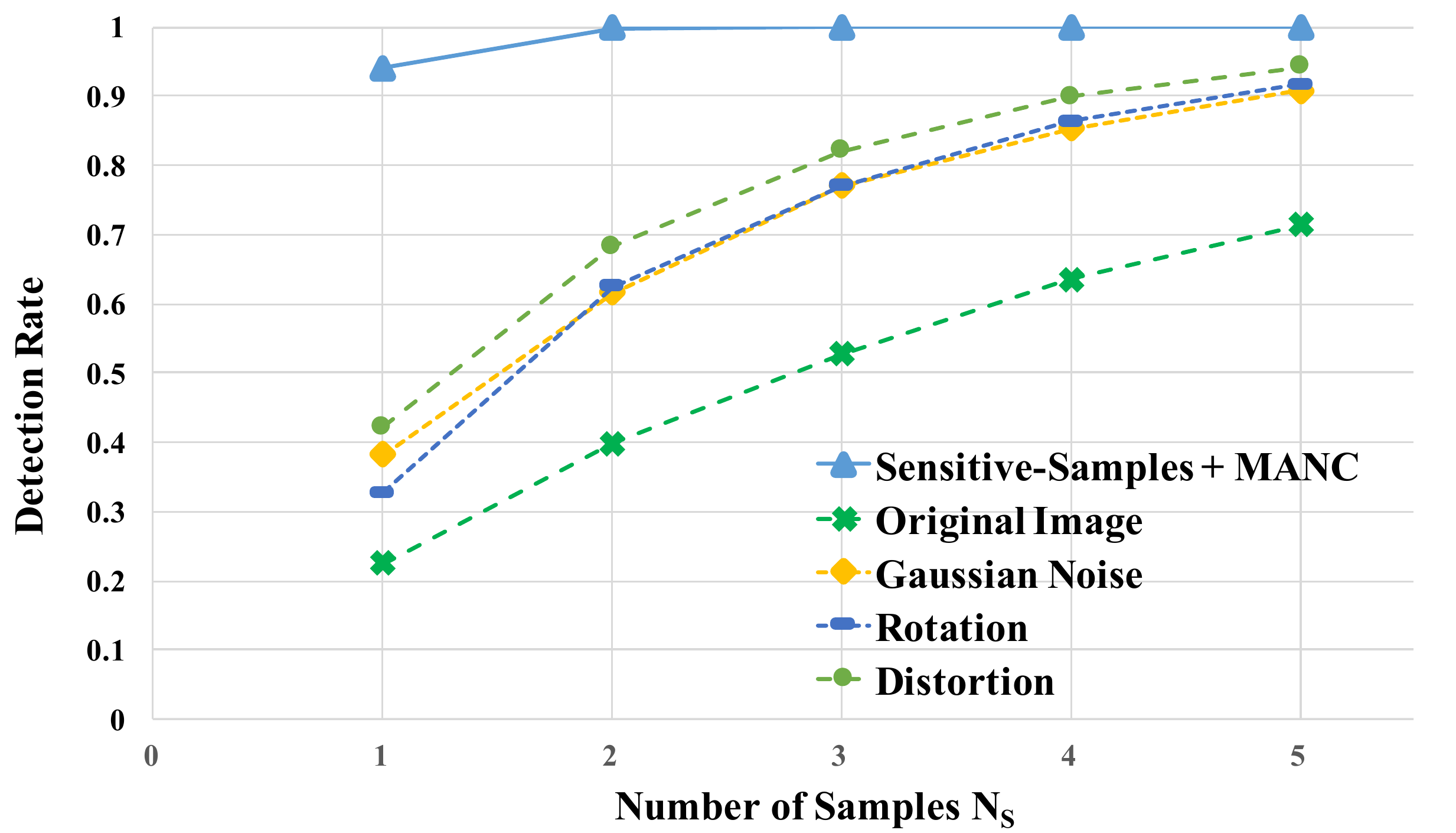}
\caption{Detection rate comparison among Sensitive-Samples and image noising, rotation and distortion. Our proposed Sensitive-Sample fingerprinting (uppermost solid blue line) achieves the highest detection rate, regardless of $N_S$.}
\label{fig:benchmark:detection-rate}
\end{figure}

We show the detection rate of different transforms in Figure \ref{fig:benchmark:detection-rate}. Our proposed \texttt{Sensitive-Sample} fingerprinting (uppermost solid blue line) achieves the highest detection rate, regardless of $N_S$. The image transforms in our consideration have higher detection rates than the original images, but not as high as our proposed solution. Considering Figures \ref{fig:benchmark:imgs} and \ref{fig:benchmark:detection-rate}, \texttt{Sensitive-Samples} are the optimal solution for model integrity verification.

\subsection{Defense Scope and Limitation}

Our proposed defense is designed for verifying the integrity of DNN \emph{models} in remote clouds. We do not consider the integrity of DNN \emph{executions} by the cloud provider. Thus defeating an adversary who compromises the execution path of DNN applications is out of our scope. There are several reasons for our defense scope. First, a lot of generic methods and systems have been proposed to protect the integrity of code execution in clouds, and these work can be applied to the deep learning inference applications. For instance, Intel SGX secure enclave is proposed to protect the executions of critical applications \cite{mckeen2013innovative} and it provides the attestation mechanism to check the security status of applications at runtime \cite{anati2013innovative}. Different approaches were designed to protect the integrity of the cloud servers \cite{McLiQu:10, AniWa:10, WaJi:10}, or cloud management services \cite{SuPeJa:14, SuPeJa:15, SzSrSe:16}. Instead, protecting the model integrity is more challenging as the adversary can tamper with the model in different ways, even before the model is transmitted to the cloud side. Second, we have described that the adversary has a variety of techniques to compromise the model without being noticed by the customer (Section \ref{sec:ps}). These subtle attack techniques are unique to deep learning. So it is necessary to design a method to protect the model integrity by the customer in the context of MLaaS. Future work includes developing methodology to verify the integrity of model inference in machine learning cloud services. 


One limitation of our proposed approach is that it cannot distinguish between malicious model changes and benign model update. Therefore, whenever the customer updates the model in the cloud, he has to re-generate new \texttt{Sensitive-Sample} fingerprint for integrity verification. 
\section{Related Work}
\label{sec:related}

We describe different types of adversarial machine learning attacks and the corresponding defenses.

\subsection{Backdoors and Trojans}
\label{sec:related:integrityattack}

\bheading{Attacks.} Past work introduced different ways to inject trojans/backdoors to compromise the model integrity. Gu et al. \cite{GuDoGa:17} proposed BadNets: the attacker retrains a correct model with poisoned training data and generates the compromised model, which predicts the images with the backdoor (a specific pattern) as an arbitrary category chosen by the attacker. This paper also showed that the backdoor remains even if the trojaned model is transfer-learned to a new one. Liu et al. \cite{Trojannn} improved this attack by only selecting a small part of the weights to change. They also reconstruct the training dataset from the model so that the adversary does not need the training dataset to inject the trojan. Chen et al. \cite{ChLiLi:17} proposed another backdoor attacks: the adversary does not need to have access to the model. He can train the model from scratch with the poisoned dataset. They evaluated the technique with different triggers and datasets.

\bheading{Defenses.} To the best of our knowledge, there exists few work defending against DNN trojans/backdoors attacks. Liu et al. \cite{LiXiSr:17} proposed three defenses: (1) anomaly detection in training data. This requires the victim to obtain the poisoned training dataset, which is not realistic; (2) retraining the model to remove the trojan. Past work has shown that it is extremely difficult to entirely remove the trojan via model retraining or transfer learning \cite{GuDoGa:17}; (3) preprocessing the input data to remove the trigger. This method forces the adversary to use the preprocessing module, which is unrealistic. Liu et al. \cite{Trojannn} proposed a solution to defeat the trojans designed by themselves: detecting the trojaned model by analyzing the distribution of mis-labeled data. However, the victim needs to feed the model with a large number of samples, making it inefficient, expensive and easily spotted when the model is served in the cloud. In contrast, our method only needs $<$10 testing samples for verification, which can be very fast and low-cost. Furthermore, our proposed method guarantees zero false-positives while the previous work does not.

\subsection{Evasion Attacks}
\label{sec:adversarysamples}

\bheading{Attacks.} Szegedy et al. \cite{SzZaSSu:13} first proposed the concept of adversarial examples: with imperceptible and human unnoticeable modifications to the input, the model mis-classifies these data. Then Goodfellow et al. \cite{GoShSz:14} proposed a fast gradient sign method to improve the speed of generating adversarial examples. Papernot et al. \cite{PaMcGo:16, PaMcGo:17} demonstrated generating adversarial examples with only black-box access to the target model. Moosavi-Dezfooli et al. \cite{NoFaFa:17} designed universal adversarial perturbations, a general perturbation to attack different images across different model architectures. These evasion attack techniques have been demonstrated in physical scenarios \cite{KuGoBe:16, EvEyFe:17} and applications, \eg, face recognition \cite{ShBhBa:16}, speech recognition \cite{CaMiVa:16, ZhYaJiTi:17}.

\bheading{Defenses.} Papernot et al. \cite{PaMcWu:16} used distillation in the training to defeat the evasion attacks. Meng and Chen \cite{MeCh:17} proposed a framework to defeat blackbox/greybox evasion attacks by detecting adversarial samples. Xu et al. \cite{XuEvQi:17} proposed a method to reduce the color bit depth or smooth the image to prevent evasion attack. These methods cannot be applied to DNN model integrity verification, as the evasion attacks do not change the parameters of the model itself.

\subsection{Poisoning Attacks}
\label{sec:PoisonAttack}

\bheading{Attacks.} An attacker can poison the training data by injecting carefully designed samples to compromise the model training process. Different poisoning techniques have been proposed for different machine learning models, \eg, SVM \cite{BiFuRo:12}, lasso regression \cite{XiBiBr:15}, topic modeling \cite{MeZh:15}, AutoRegressigve models \cite{AlZhBa:16}. Gradient-based poisoning attacks were designed against deep neural networks \cite{MuBiDe:17, YaWuLi:17, KoLi:17}. 

\bheading{Defenses.} Defense solutions have been proposed to defeat poisoning attacks. The most popular method is to detect and remove the poisoning data from the dataset by statistical comparisons \cite{ChStVa:17, LiLiVo:17, StKoLi:17}. Poisoning attacks compromise the model and make it misbehave under normal input, thus degrading the performance of the learning system. In contrast, the adversary in DNN trojan attacks may only inject a small number of poisoned data into the training dataset, to make the model behave abnormally only with targeted inputs.
\section{Conclusion}
\label{sec:conclu}

As deep learning cloud services become popular, protecting the integrity of DNN models becomes urgent and important. In this paper, we show that the integrity of model weights can be dynamically verified by just querying the deployed model with a few transformed inputs and observing their outputs. Our proposed detection method defines and uses Sensitive-Samples, which introduces sensitivity of DNN outputs corresponding to the weights. Even a small modification of the model weights can be reflected in the outputs. We introduce more constraints to ensure the generated sensitive-samples are similar to the original inputs, in order to avoid being noticed by the adversary. Our evaluation on different categories of DNN integrity attacks shows that our detection mechanism can effectively ($>$99\% detection rate) and efficiently ($<$10 black-box accesses) detect DNN integrity breaches.

\bibliographystyle{IEEEtranS}
\bibliography{ref}

\end{document}